\documentclass[aip,graphicx]{revtex4-1}
\usepackage{graphicx}
\setcitestyle{numbers,square}
\usepackage{amsmath,xcolor}
\usepackage{booktabs}
\usepackage{relsize}
\draft 
\usepackage[colorlinks = true,
        urlcolor  = blue,citecolor=blue]{hyperref}

\begin{document}


\title{Stringlet Excitation Model of the Boson Peak} 



\author{Cunyuan Jiang}
\affiliation{School of Physics and Astronomy, Shanghai Jiao Tong University, 200240, Shanghai, China}
\affiliation{Wilczek Quantum Center, Shanghai Jiao Tong University, 200240, Shanghai, China}
\affiliation{Shanghai Research Center for Quantum Sciences, 200240, Shanghai, China}
\author{Matteo Baggioli}
\email[]{b.matteo@sjtu.edu.cn}
\affiliation{School of Physics and Astronomy, Shanghai Jiao Tong University, 200240, Shanghai, China}
\affiliation{Wilczek Quantum Center, Shanghai Jiao Tong University, 200240, Shanghai, China}
\affiliation{Shanghai Research Center for Quantum Sciences, 200240, Shanghai, China}
\author{Jack F. Douglas}
\affiliation{Materials Science and Engineering Division,National Institute of Standards and Technology, 20899, Gaithersburg, Maryland, United States}


\date{\today}

\begin{abstract}
The boson peak (BP), a low-energy excess in the vibrational density of states over the Debye contribution, is often identified as a characteristic of amorphous solid materials. Despite decades of efforts, its microscopic origin still remains a mystery. Recently, it has been proposed, and corroborated with simulations, that the BP might stem from intrinsic localized modes involving one-dimensional (1D) string-like excitations (``stringlets"). We build on a theory originally proposed by Lund that describes the localized modes as 1D vibrating strings, but we specify the stringlet size distribution to be exponential, as observed in simulations. We provide an analytical prediction for the BP frequency $\omega_{BP}$ in the temperature regime well below the observed glass transition temperature $T_g$. The prediction involves no free parameters and accords quantitatively with prior simulation observations in 2D and 3D model glasses based on inverse power law potentials. The comparison of the string model to observations is more uncertain when compared to simulations of an Al-Sm metallic glass material at temperatures well above $T_g$. Nonetheless, our stringlet model of the BP naturally reproduces the softening of the BP frequency upon heating and offers an analytical explanation for the experimentally observed scaling with the shear modulus in the glass state and changes in this scaling in simulations of glass-forming liquids. 

Finally, the theoretical analysis highlights the existence of a strong damping for the stringlet modes above $T_g$, which leads to a large low-frequency contribution to the 3D vibrational density of states, observed in both experiments and simulations.
\end{abstract}


\maketitle 

\section{Introduction}
Amorphous solids exhibit a variety of characteristic anomalies in their vibrational, thermodynamic and transport properties, when compared to the more familiar case of ordered crystalline matter \cite{PhysRevB.4.2029,doi:10.1142/q0371}. Among the different, and apparently universal \cite{ramos2020universal}, anomalous properties of glassy materials, the boson peak (BP) is probably the most commonly discussed and controversial. This peak refers to an excess in the vibrational density of states (VDOS) normalized by the phonon density of states of Debye theory, $g(\omega)/\omega^{d-1}$, where $d$ is the number of spatial dimensions and $\omega$ the frequency. The BP is ubiquitously observed in amorphous systems \cite{ALEXANDER199865} and recently this feature has also been observed in crystalline systems at finite temperature (\textit{e.g.}, \cite{RevModPhys.86.669,PhysRevB.99.024301}), so it is unclear whether it arises from structural disorder \cite{PhysRevLett.122.145501}.

Several theoretical models have been proposed to rationalize the BP anomaly in glasses \cite{Elliott_1992,Schirmacher_2006,PhysRevLett.98.025501,PhysRevLett.115.015901,PhysRevLett.97.055501,PhysRevB.43.5039,PhysRevB.67.094203,PhysRevB.76.064206,PhysRevE.61.587,PhysRevLett.122.145501,PhysRevLett.86.1255,PhysRevLett.106.225501,doi:10.1080/00018738900101162,KLINGER2002311,Grigera2003}, but a final verdict has not yet been reached. A common idea is that the BP represents a signature for the emergence of additional vibrational modes coexisting, but distinct from phonons. These ``excess modes" have been suggested to arise from both structural disorder or from the anharmonicity in intermolecular interactions inherent to the liquid state and heated crystals. These commonly observed vibrational modes are expected to be effectively localized as they involve a rather limited number of atoms or molecular segments in complex molecules \cite{PhysRevB.67.094203,PhysRevB.76.064206,PhysRevB.53.11469,2022arXiv221010326L,10.1063/5.0147889,2023arXiv230403661M,10.21468/SciPostPhysCore.4.2.008,VURAL20113528,SCHOBER1993965}. At large enough wave-vector, they also interact with phonons, making them overdamped, \textit{i.e.}, reaching the Ioffe-Regel (IR) limit \cite{Shintani2008,PhysRevB.87.134203,PhysRevLett.96.045502}, leading to hindered thermal transport because of the consequent strong phonon scattering.

\begin{figure}
    \centering
    \includegraphics[width=0.48\linewidth]{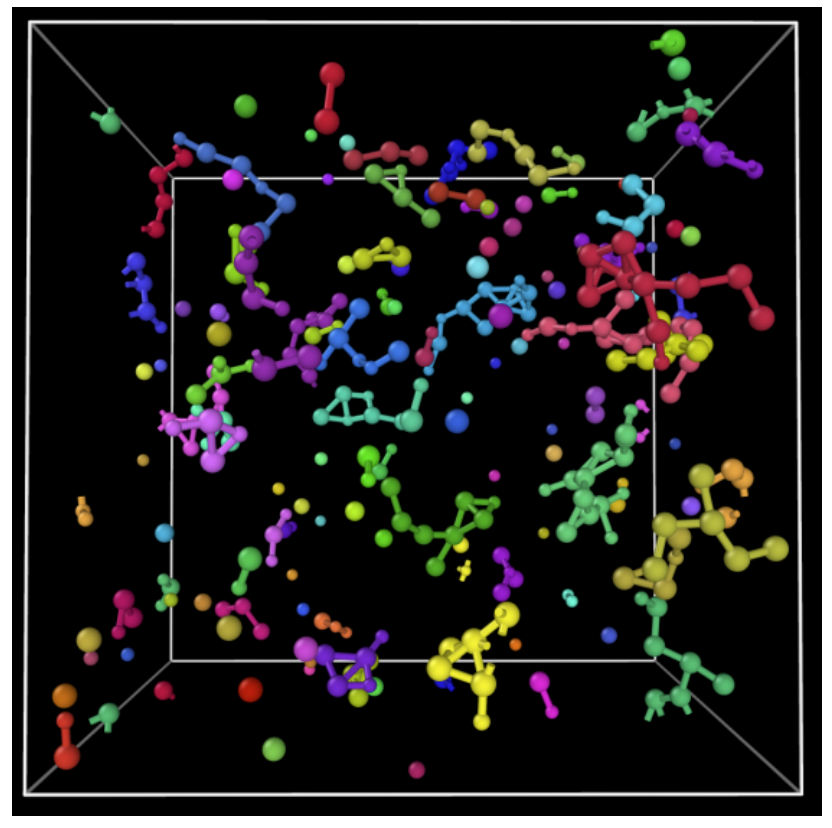}\quad
     \includegraphics[width=0.48\linewidth]{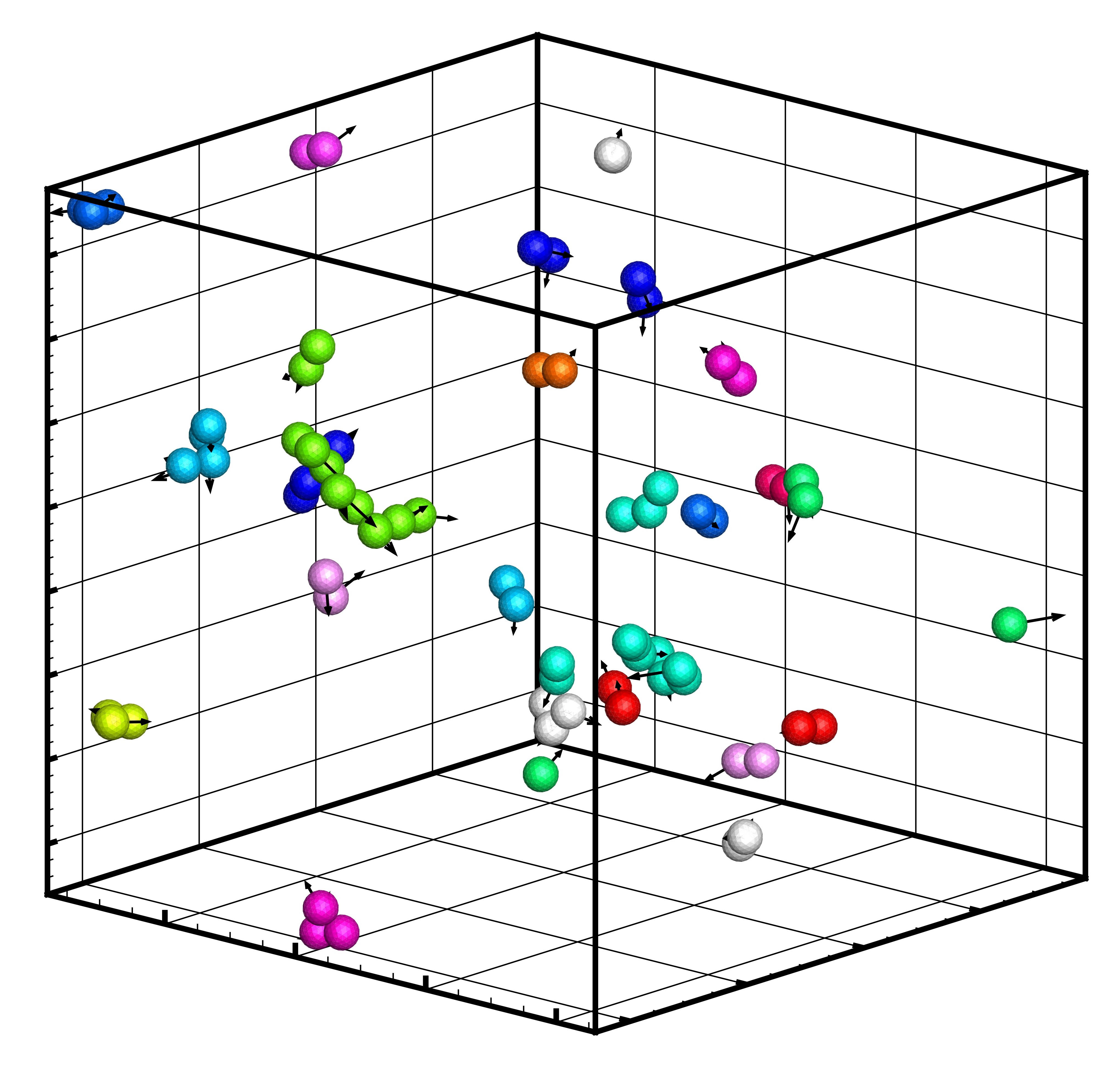}
    \caption{\textbf{Left panel: }A concrete visualization of string-like (``stringlets'') excitations involving cooperative particle exchange motion, and corresponding to localized modes having a frequency around the BP value. Figure taken with permission from \cite{PhysRevResearch.5.023055}. \textbf{Right panel: } A stringlet atomic configuration from the simulation Al-Sm metallic glass system \cite{10.1063/5.0039162} at $550$K.}
    \label{fig:0}
\end{figure}

Despite numerous experimental and simulation studies, a fully predictive theoretical model explaining the existence of these modes
and how they give rise to the observed boson peak remains elusive. Recently, it has been recognized that, on atomic scales, localized string-like excitations that involve reversible particle exchange motion in the form of linear polymeric structures exist in glass-forming liquids. These dynamic structures have been termed ``\textit{stringlets}'' \cite{10.1063/5.0039162}. See Fig.\ref{fig:0} for a visual representation from simulations. Moreover, it has been shown \cite{C2SM27533C,10.1063/5.0039162} that these stringlets are the dominant contribution to the boson peak and grow upon heating, leading to the material softening upon heating. A similar phenomenon was originally observed in the glassy interfacial dynamics of crystalline Ni nanoparticles in their ``premelted" state, where anharmonic interactions in the crystalline material are prevalent \cite{C2SM27533C,C2SM26789F}. The existence of these structures and their significance for the boson peak have been confirmed in both 2D and 3D amorphous systems by Hu and Tanaka \cite{Hu2022,PhysRevResearch.5.023055}. Moreover, Betancourt et al. \cite{10.1063/1.5009442,doi:10.1073/pnas.1418654112} have shown a direct relation between the stringlets and string-like collective motion that has often been observed in connection with activated transport occurring on a much longer timescale. A general relation between the stringlet size and the mean square displacement on the fast $\beta$-relaxation time, a quantity that has been related to the long time $\alpha$-relaxation time, has been observed as well \cite{Douglas_2016}. Recently, Hu and Tanaka confirmed and extended these relations between the fast dynamics and $\alpha$-relaxation \cite{Hu2022,PhysRevResearch.5.023055}. Experimental studies have long suggested that transverse modes might give a predominant contribution to the boson peak \cite{PhysRevB.12.2432,TScopigno_2003,tian2021structural} and later simulation studies have provided strong support for this proposition \cite{Schober_2004,Shintani2008}. Additionally, it has also been noted that excess modes of a one-dimensional nature \cite{novikov1990spectrum} are consistent the commonly observed near linear frequency dependence of the light-vibrational coupling parameter $C(\omega)$ \cite{achibat1993correlation,PhysRevB.48.7692,Surovtsev_2004}. These observations and their interpretation are broadly in accord with the idea that topological ``defects'' can give rise to the BP in glasses \cite{Angell_2004,BHAT20064517}, and also with the related suggestion that these excitations are important for the plasticity and yield strength of glassy materials \cite{PhysRevLett.127.015501,Baggioli2023,PhysRevE.105.024602,Wu2023}.

We note that previous models \cite{MALINOVSKY199163,Hu2022,PhysRevResearch.5.023055} phenomenologically assumed that the excess modes responsible for the boson peak involve structures with a log-normal size distribution, but the physical basis of this assumption has been completely unclear. Lund \cite{PhysRevB.91.094102} recently sought to describe these excess normal modes using a continuum solid model in which a random distribution of line ``defects'' or ``elastic strings" was assumed to exist based on simulation observations. Lund's model is an extension of the elastic string model of Granato and L\"{u}cke \cite{granato1956theory}, introduced long ago in the description of the plasticity in highly defective crystals, where the strings correspond to pinned segments of dislocation lines in an extended dislocation network. We note that an even earlier string model of the boson peak was proposed by Novikov \cite{novikov1990spectrum}. Lund's model \cite{PhysRevB.91.094102} was able to rationalize many of the anomalous features of glasses, including the presence of a BP anomaly, Rayleigh sound attenuation constant, a negative sound dispersion for acoustic phonons with a minimum near the BP, and the saturation of the IR limit for transverse phonons. Nevertheless, the application of this model \cite{PhysRevB.101.174311} to estimate the length distribution $p(l)$ of these hypothetical string-like structures, based on experimental observations of the VDOS in glycerol and silica, unfortunately led to a string length distribution inconsistent with simulation observations \cite{10.1063/5.0039162,Hu2022,PhysRevResearch.5.023055}, which appears to instead indicate an approximate exponential form, $p(l)\sim \exp(- l/\lambda)$. We address this inconsistency below. Before continuing, we emphasize that despite several commonalities, stringlets are strictly speaking not ``structural defects'' as found in crystalline materials, even if they share some geometric similarities to dislocation networks. In this context, ``string-like” dislocation segments exist within the dislocation loop networks where the points of impingement of the dislocation loops define pinning points or end-points of the dislocation segments \cite{granato1956theory,PhysRevB.72.174110}. As we shall discuss extensively below, string-like structures generally arise also in glass-forming liquids, which apparently have the same length distribution assumed in the modeling of dislocation segments. This situation allows us to model these structures based on a similar coarse-grained string mathematical model. We emphasize, however, that the stringlets of glass-forming liquids should not be equated with dislocation segments arising in mechanically deformed crystalline materials. A previous discussion of this important point can be found in Section V of Ref.\cite{PhysRevB.91.094102}, and we discuss this ``analogy" further in Appendix \ref{app1}. Moreover, this form of collective motion on the timescale of the fast beta relaxation process, occurring generally on a timescale on order of a ps, should not be identified with collective motion associated with thermal activation events associated with the alpha relaxation process that occurs on a much longer timescale associated with the average rate of intermittent hopping events associated with particle diffusion and the structural relaxation of the material as a whole \cite{10.1063/1.5009442,10.1063/1.4878502}. As we shall discuss below, Hong et al. \cite{PhysRevE.83.061508,HONG2011351} have heuristically identified the boson peak with this alternative form of collective motion.

Here, we revisit the Lund string model of the boson peak \cite{PhysRevB.91.094102} and extend it to make a parameter-free prediction of $\omega_{BP}$ in both 2D and 3D for amorphous solids. In contrast to the work of Lund, however,
we assume an exponential form for the string length distribution $p(l)$, at the outset to be consistent with simulation observations \cite{10.1063/5.0039162,Hu2022,PhysRevResearch.5.023055}, and we further assume that the average stringlet vibration speed equals the transverse sound velocity to uniquely fix our model parameters. This model leads an extremely simple analytic expression for the BP frequency without any fitting parameters, which appears to be highly predictive in the the low temperature glass state, where many boson peak measurements are normally performed. In particular, our predictions are in good agreement with simulation results for the BP frequency at $T=0$, with an uncertainty between $1\,\%$ and $5\,\%$. The stringlet model also captures the correct qualitative trend at high temperatures where these vibrational modes begin to become overdamped. Interestingly, we are also able to reproduce the softening of the BP frequency with temperature and to explain the experimentally observed scaling of $\omega_{BP}$ with the shear modulus $G$ \cite{Tomoshige2019}. Our theoretical model of the boson peak suggests that a strong damping of the stringlet modes is required  above $T_g$ to describe simulation observations. As a by-product, the introduction of a friction term for the string vibrations produces a low-frequency divergence in the VDOS normalized to the Debye law, $g(\omega)/\omega^{d-1}$, which is ubiquitously observed in experimental and simulation data at finite temperature \cite{PhysRevE.83.061508,BUCHENAU1993275} (resembling similar features in anharmonic crystals and incommensurate structures), and usually not discussed in most of the traditional theories for the BP (see \cite{PhysRevLett.93.245902,Baggioli_2020,PhysRevResearch.1.012010,doi:10.1142/S0217979221300024,jiang2023glassy} for some exceptions).\\

\section{The stringlet model}
We initially directly follow Lund et al. \cite{PhysRevB.91.094102,PhysRevB.101.174311}, and model the stringlets as 1D vibrating strings that are anchored at both ends. Similar results follow by assuming that the strings are free at both ends, since, in both cases, the vibrations would be characterized by a fundamental wave-vector $k=\pi/l$, with $l$ the length of the string. As noted before, 1D vibrational modes being excited by light are consistent with the low-frequency Raman scattering observations on glasses, where a near linear scaling for the light to vibration coupling coefficient $C(\omega)$ has been found over a large frequency range \cite{PhysRevB.59.38}. According to this simple continuum based model, which is agnostic to the actual physical origin of these vibrational modes, each stringlet of length $l$ has a corresponding frequency of vibration given by,
\begin{equation}
    \omega_l=\sqrt{\left(\frac{\pi v}{l}\right)^2-\gamma^2}\equiv \sqrt{\omega_0^2-\gamma^2}\label{ab}
\end{equation}
where $v$ is the speed of propagation of the stringlet wave, and $\gamma$ represents a ``friction parameter'' (see \cite{PhysRevB.91.094102,PhysRevB.101.174311} for details). Eq.\eqref{ab} is based on the assumption that only the fundamental frequency of vibration of the stringlet is relevant, \textit{i.e.}, higher harmonics are neglected. 
We consider a set of stringlets whose length follows a distribution denoted as $p(l)$. Based on this assumption, the stringlet density of states $g_s(\omega)$ can readily be obtained using the \textit{inverse distribution function}. More precisely, given the distribution function of a random variable $X>0$, the distribution function of the inverse variable $Y\equiv 1/X$ can be derived by relating the cumulative distribution function of the initial variable $F(x)$ to that of the inverse variable,
\begin{equation}
  G(y)=1-F\left(\frac{1}{y}\right).
\end{equation}
The density function of the inverse variable $Y$ can be then obtained as the derivative of the cumulative distribution function:
\begin{equation}
    g(y)=\frac{1}{y^2}f\left(\frac{1}{y}\right).\label{corr}
\end{equation}
Following this procedure, the frequency distribution of the stringlets can be obtained as,
\begin{equation}
    g_s(\omega_l) \,d\omega_l=-(d-1)\, p(l)\, dl \label{ide}
\end{equation}
where $d$ is the spatial dimension. Notice how the minus sign in the above equation, which can be formally derived using the procedure described above, ensuring the physically required positivity of the stringlet density of states $g_s(\omega)$ presented below. In three dimensions ($d=3$), the factor of $d-1=2$ in the r.h.s. of Eq.\eqref{ide} reflects the ability of the stringlets to oscillate along two linearly independent directions that are perpendicular to its equilibrium orientation. Formally, this framework is equivalent to considering a distribution of damped oscillators with frequency $\omega_0$ and stringlet damping rate $\gamma$. We can then define the quality factor $Q=\omega_0/\gamma$ according to which $Q\ll 1$ corresponds to underdamped vibrations and $Q$ of $\mathcal{O}(1)$ to overdamped ones. Notice that a priori $Q$ depends on the stringlet length $l$ since $\omega_0$ exhibits this property.

We next introduce a new assumption -- with respect to previous work \cite{PhysRevB.91.094102,PhysRevB.101.174311} -- into this type of quasi-continuum framework that is an exponential distribution for the stringlet length,
\begin{equation}
    p(l)=p_0 \,e^{-l/\lambda}\label{di}
\end{equation}
which is motivated directly by simulation observations \cite{10.1063/5.0039162,Hu2022,PhysRevResearch.5.023055}. The prefactor in Eq. \eqref{di} is determined by the distribution normalization. Notice that this distribution is different from what assumed in Ref. \cite{PhysRevB.101.174311}, where a localized Gaussian-like distribution of lengths is considered (see Fig.7 therein). As shown in Refs. \cite{10.1063/5.0039162,Hu2022,PhysRevResearch.5.023055}, this assumption is inconsistent with simulation data. 

The above distribution, Eq.\eqref{di}, presents a natural lower cutoff $l_0$ that corresponds to the size of the smallest particle in the system. In particular, this implies the consistency condition $l>l_0$, \textit{i.e.}, physically a stringlet has to be larger in size than the size of an individual particle. The average stringlet length $\langle l \rangle$ imay be defined as,
\begin{equation}
    \langle l \rangle = \frac{\int_{l_0}^\infty l \,p(l) dl}{\int_{l_0}^\infty p(l) dl}=l_0+\lambda.
\end{equation}
When expressed in units of the smaller particle diameter, as done below, this relation reduces to $\langle l \rangle = 1+\lambda$.

Combining Eq.\eqref{ab} with Eq.\eqref{ide}, we obtain obtain a closed form analytic expression for the stringlet VDOS,
\begin{equation}
    g_s(\omega)=p_0\,\pi (d-1) \, v\,\frac{ \omega \,e^{-\frac{\pi  v}{\lambda\,\sqrt{\gamma^2
   +\omega^2}}}}{\left(\gamma^2 +\omega^2\right)^{3/2}}\,.\label{final}
\end{equation}
Because of our current lack of understanding of the normalization of the stringlet density of states, and thus the total number of stringlets, we add an unknown constant in front of the above expression that takes into account the normalization of the stringlet density of states. This constant can only be determined by means of a microscopic theory of the stringlets or can otherwise be taken as a fitting parameter from the data, as done in our case. We therefore find it convenient to rewrite the above expression as,
\begin{equation}
    g_s(\omega)=\mathcal{N}_s \,\frac{ \omega \,e^{-\frac{\pi  v}{\lambda\,\sqrt{\gamma^2
   +\omega^2}}}}{\left(\gamma^2 +\omega^2\right)^{3/2}}\,,\label{final2}
\end{equation}
where all the prefactors have been absorbed into the prefactor $\mathcal{N}_s$, an empirical parameter since its determination requires knowledge of the total number of stringlets. Before continuing, we notice that the value of $\mathcal{N}_s$ does not affect the prediction of the BP frequency but this quantity does influence its intensity. As one might expect, the stringlet model of the boson peak predicts that the larger the number of stringlets the stronger the intensity of the BP. We note that simulation could provide information about the number of stringlets, but no information of this kind has so far been collected in simulation studies. It would be particularly interesting to determine also if the number of stringlets correlates with the fragility of glass-formation.

In the limit of zero stringlet damping, a situation that applies physically well below $T_g$, Eq. \eqref{final2} reduces to the particularly simple form,
\begin{equation}
    \lim_{\gamma \rightarrow 0}g_s(\omega)=\mathcal{N}_s\,\frac{  \,e^{-\frac{\pi   v}{\lambda\,\omega}}}{\omega^2}\,.\label{final3}
\end{equation}
From Eq.\eqref{final2}, we can see that in two spatial dimensions the Debye normalized stringlet VDOS $g_s(\omega)/\omega$ has a maximum $\omega_{max}$,
\begin{equation}\label{twotwo}
    \omega_{max}^{2D}=\frac{1}{3} \sqrt{\pi ^2 v^2/\lambda^2-9 \gamma^2 },
\end{equation}
which in the limit of zero damping becomes simply,
\begin{equation}\label{w1}
    \omega_{max}^{2D}(\gamma=0)=\frac{\pi}{3}\,\frac{v}{\lambda}.
\end{equation}
In 3D, there is also an analytical maximum in $g_s(\omega)/\omega^2$, but the resulting analytic expression is somewhat complicated. The effect of the friction $\gamma$ is always that of softening the bare prediction, as shown explicitly for the 2D case in Eq.\eqref{twotwo}. The 3D solution simplifies in the zero damping limit, for which we obtain the concise result,
\begin{equation}\label{w2}
    \omega_{max}^{3D}(\gamma=0)=\frac{\pi}{4}\,\frac{v}{\lambda}.
\end{equation}
More generally, as discussed previously in \cite{10.1063/1.4878502}, and below, the stringlet distribution is not a perfect exponential, as expected from mean field theories of linear string formation at equilibrium. A better, more general, representation involves a combination of an exponential form with a power-law prefactor \cite{10.1063/1.4878502},
\begin{equation}\label{corr}
    p(l)= p_0 \,l^{-\theta}\,e^{-l/\lambda},
\end{equation}
where the exponent $\theta$ depends on the topology of the string-like structures, and their excluded volume and other interactions.
It is straightforward to re-derive the stringlet VDOS using this distribution. Again, we absorb constants into a modified prefactor, $\mathcal{\tilde{N}}_s$. This yields the relation,
\begin{equation}
 \tilde   g_s(\omega)= \mathcal{\tilde{N}}_s \,\omega \left(\gamma ^2+\omega^2\right)^{(\theta-3)/2 } e^{-\frac{\pi  v}{\lambda  \sqrt{\gamma ^2+\omega^2}}}
\end{equation}
which reduces to Eq.\eqref{final2} when $\theta=0$. For zero damping, $\gamma=0$, this is simply,
\begin{equation}
    \lim_{\gamma\rightarrow 0} \tilde g_s(\omega)=\mathcal{\tilde{N}}_s\, \omega^{\theta -2} e^{-\frac{\pi  v}{\lambda  \omega}}.
\end{equation}
In 3D, and in the limit of zero damping, the reduced stringlet VDOS $\tilde g_s(\omega)/\omega^2$, displays a maximum set by the condition,
\begin{equation}
   \tilde  \omega_{max}^{3D}(\gamma=0)=\frac{\pi}{(4-\theta)}\,\frac{v}{\lambda},
\end{equation}
which is a simple generalization of Eq.\eqref{w2}.

Finally, we consider the full vibrational density of states $g_{\mathrm{total}}(\omega)$, including the phononic Debye contribution. We then arrive at the simple relation,
\begin{equation}
    g_{\mathrm{total}}(\omega)=g_{\text{Debye}}(\omega)+g_s(\omega),\label{sum}
\end{equation}
where, in 3D for example,
\begin{equation}
    g_{\text{Debye}}(\omega)=\frac{\omega^2}{2\pi^2 \bar{v}^3},\quad \text{and}\quad \frac{1}{\bar{v}^3}=\frac{1}{v_L^3}+\frac{2}{v_T^3},
\end{equation}
where $v_L,v_T$ are respectively the speed of longitudinal and tranverse acoustic phonons. The arguments above indicate that the boson peak frequency $\omega_{BP}$ coincides with the location of the maximum in the stringlet contribution to the VDOS, $g_s(\omega)$. The independence of the two contributions to the total density of states in Eq.\eqref{sum} relies on the simplifying assumption that phonons and stringlets are non-interacting. As we will see below, this simple expression is sufficient to capture the most salient features related to the boson peak. It is certainly possible to go beyond this leading order approximation by considering the interactions between stringlets and phonons. Formally, this might be done by modifying the respective dynamical equations, in a way similar to what done for dislocations and phonons in ordered crystals (\textit{e.g.}, see Refs. \cite{PhysRevB.72.174110,PhysRevB.72.174111}). While this type of coupling is naturally expected to exist, our results below indicate that, at least in the low temperature limit, it is not fundamental in deriving the position of the BP frequency. Nevertheless, we expect such coupling to be crucial to describe other properties in amorphous solids such as sound attenuation and thermal conductivity.

We notice that a similar expression for the BP frequency as in Eqs.\eqref{w1}-\eqref{w2} was suggested before by Granato \cite{PhysRevLett.68.974}, Hong et al. \cite{PhysRevE.83.061508,HONG2011351} and Kalampounias et al. \cite{10.1063/1.2360275} on a more heuristic basis (See also Refs.\cite{MALINOVSKY1988111,MALINOVSKY1986757} and Ref.\cite{https://doi.org/10.1002/pssb.2220640120} for the original idea of introducing a length scale associated with the boson peak.). The length-scale appearing in Eqs.\eqref{w1}-\eqref{w2} has been interpreted by Hong et al.\cite{PhysRevE.83.061508,HONG2011351} in recent work as a ``cooperativity length scale'' $\xi$ associated with cooperative motion associated with the growing activation energy of the alpha relaxation time of glass-forming liquids postulated in the Adam-Gibbs model \cite{10.1063/1.1696442}. However, there is some evidence that this characteristic length defined from an equation of the assumed form of Eqs.\eqref{w1} and \eqref{w2} increases with temperature, as observed for the stringlet length in simulations \cite{10.1063/5.0039162,C2SM26789F}. In contrast, the activation and the associated scale of collective motion in glass-forming liquids grows upon cooling, Refs.\cite{10.1063/1.5009442,10.1063/1.4878502}, a trend that is contrary to the intuitively attractive idea of identifying $\xi$ with the size of the ``cooperatively rearranging regions'' of Adam-Gibbs. Independent of these experimental observations on the boson peak, string-like collective motion whose extent grows with temperature has been inferred from low angle inelastic coherent neutron scattering measurements \cite{PhysRevLett.84.3630}. Notice that the numerical prefactor in Eqs.\eqref{w1}-\eqref{w2}, defining the characteristic length associated with the boson peak, is completely unspecified in previous studies, while it is specified exactly in our model. We finally remark that our theoretical approach presents two important improvements with respect to the earlier phenomenological studies reported above. First, the length scale appearing in the expression for the BP frequency is immediately identified with the stringlet length obtained from simulations observations. At the same time, the numerical prefactor describing the stringlet size distribution can be directly extracted from the simulation data. Compared to the previous literature, our theoretical model is based on more physical arguments that go beyond a pure dimensional analysis or vague ideas about cooperative motion of some unspecified kind. Nevertheless, it is rather remarkable that the brilliant raw intuition behind the qualitative picture of Sokolov and coworkers \cite{PhysRevE.83.061508,HONG2011351} anticipated the functional form predicted by our model.\\

In summary, this simple model exhibits three striking predictions. (I) It gives an analytical and simple estimate for the BP frequency in both 2D and 3D systems, Eqs.\eqref{w1}-\eqref{w2}. These results imply that the BP in a 3D system should appear at lower energies compared to a 2D system with the same length distribution and sound speed. (II) If we reasonably assume that the speed of the stringlet excitation is the same of that of transverse phonons, this model immediately predicts that $\omega_{BP}\sim G^{1/2}$ in the low temperature glass state where the stringlet average length
(and the configurational entropy as well) are expected to become independent of temperature \cite{doi:10.1073/pnas.1418654112,10.1063/1.5009442,Douglas_2016}. However, the stringlet model also predicts that this simple scaling becomes progressively modified at temperatures approaching $T_g$, where the temperature dependence of $\lambda$ can no longer be neglected. Importantly, identifying the velocity of propagation of stringlet excitations with the average speed of transverse sound is a working hypothesis in our coarse-grained model. Nevertheless, this hypothesis can be motivated by previous observations in simulations \cite{Shintani2008,PhysRevResearch.5.023055}  and measurements \cite{PhysRevLett.69.1540,PhysRevLett.78.2405} that the boson peak strongly correlates with the transverse sound modes. Moreover, we remark that Sokolov and coworkers \cite{PhysRevE.83.061508,HONG2011351} have recently analyzed the boson peak in diverse materials with a view of defining a dynamical correlation length for collective motion based on an assumed expression in which the boson peak is taken to be proportional to the transverse sound velocity divided by a cooperative length scale (the analog of the stringlet ``size'' in our model). (III) At highly elevated temperatures where we may expect the glass to become a liquid, however, the stringlet model interestingly predicts a linear frequency dependence of the VDOS, and thus leads to a $1/\omega$ divergence of the reduced density of states, the density of states normalized by the Debye density of states. This scaling is consistent with recent modeling of the density of states so that the stringlet model appears to recover a physically correct behavior in the high temperature simple liquid as well as in the low temperature glass states.

Before proceeding with our results, we notice that the identification and characterization of the stringlets from simulations can be done in different ways and can be achieved from the analysis of the collective dynamics on the $\beta$ relaxation time scale. This is how the stringlets were originally defined and characterized both in liquids and amorphous solids, see for example Ref. \cite{10.1063/1.5009442} (in particular figure 3 from which the stringlet length can be extracted) and Ref. \cite{PhysRevLett.80.2338} for the original method and idea. In this sense, the stringlet distribution $p(l)$ is an independent input that can be obtained from simulations (without diagonalization of the dynamical matrix) and then used to predict the BP location in the density of states. We note that the diagonalization of the dynamical matrix involves a harmonic approximation which is best to avoid if possible. Tanaka and collaborators \cite{Hu2022}, more recently, introduced a new method to identify and characterize these stringlets by looking at the particle-level vibrational density of states. Regarding this second method, the identification of the stringlets with the boson peak necessitates a knowledge of the density of states and anomalous excess modes from those expected from Debye theory. There are a variety of methods of determining the density of states, however.

\section*{Prediction of the boson peak deep in the glass state}
We next consider the capacity of the revised string model of the boson peak to quantitatively account for the frequency dependence of the boson peak. First, we obtain the value of the parameter $\lambda$ in Eq.\eqref{di} by fitting the distribution of stringlet length obtained from simulations. Then, we take the value for the velocity of transverse phonons $v_T$ which is obtained by fitting the dispersion relation obtained from simulations. Once these two parameters are known, and fixed by the simulations, our theoretical model provides a parameter-free prediction for the BP frequency in both 2D and 3D amorphous solids, Eqs.\eqref{w1}-\eqref{w2}. We can then directly compare the predicted value for $\omega_{BP}$, which we will denote as $\omega_{BP}^{th}$, with the numbers obtained from the simulation data (as reported in the original references), which we will label as $\omega_{BP}^{\mathrm{sim}}$.\\

\begin{figure}
    \centering
    \includegraphics[width=0.45\linewidth]{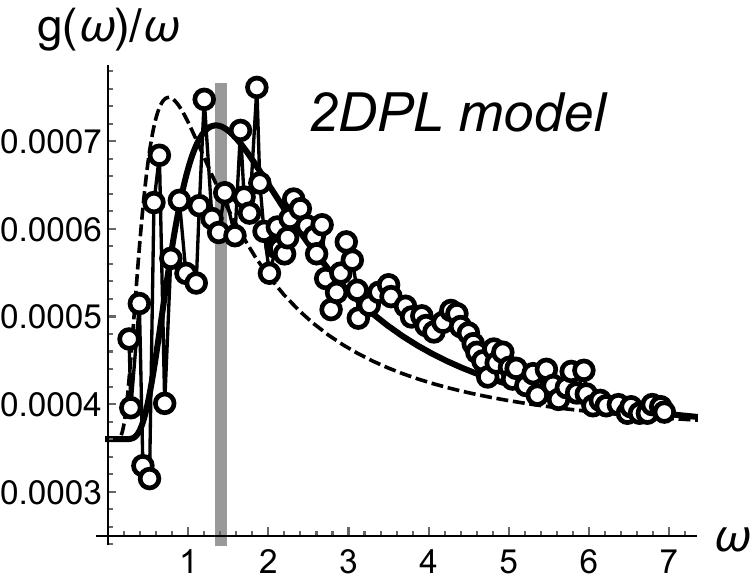} \qquad    \includegraphics[width=0.45\linewidth]{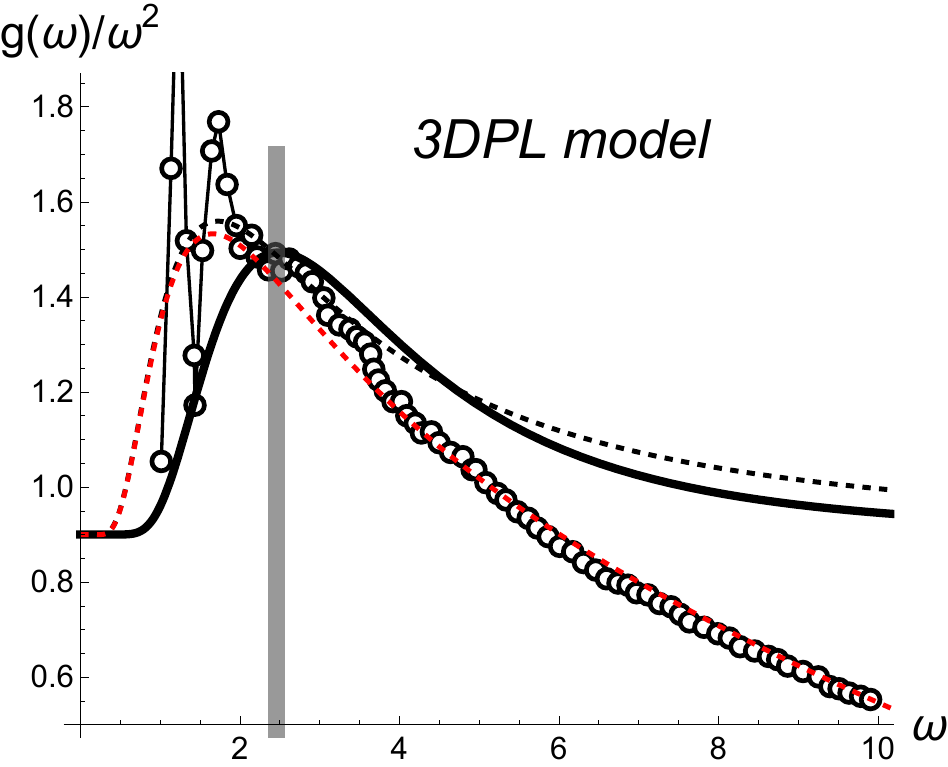} \\

    \vspace{0.5cm}

    \includegraphics[width=0.45\linewidth]{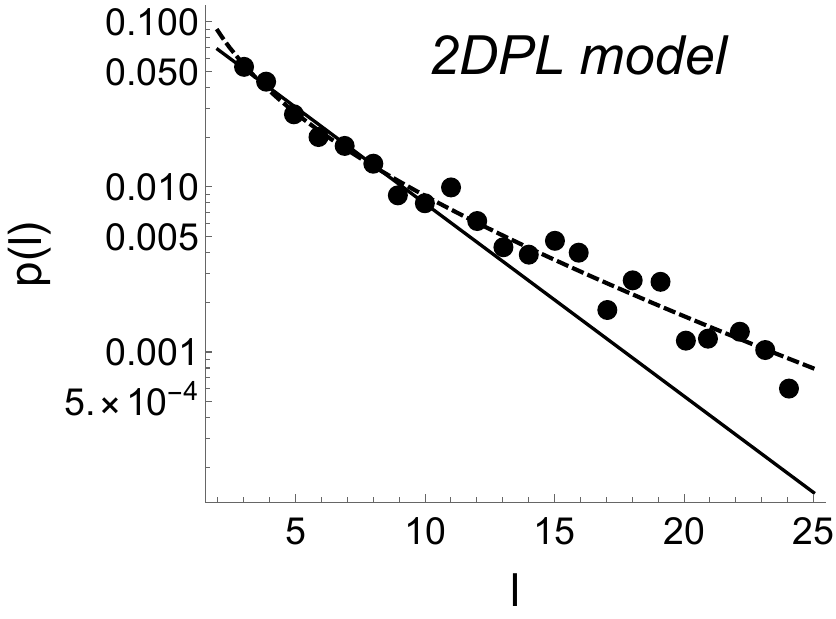} \qquad    \includegraphics[width=0.45\linewidth]{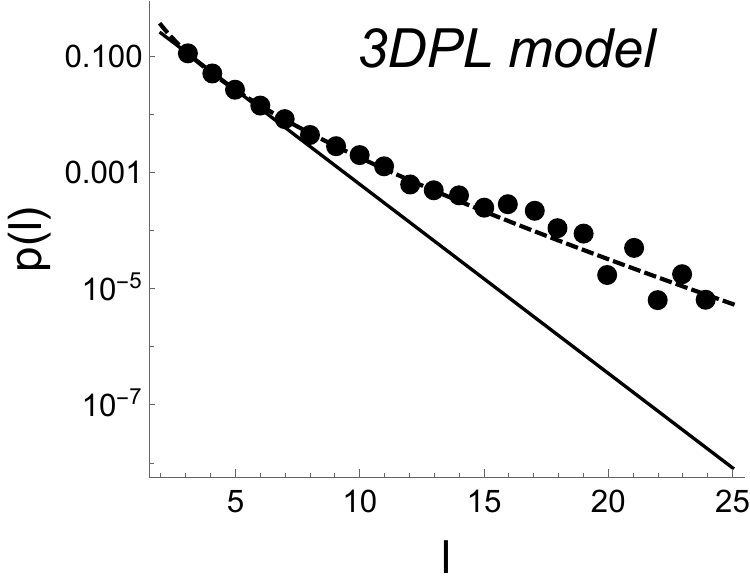} 
    \caption{A comparison between the theoretical predictions and simulation data for the reduced density of states. \textbf{(Top):} The normalized density of states for the 2D and 3D systems. The solid black lines are the theoretical predictions using the original exponential distribution for the stringlet size. The black dashed lines are the theoretical predictions with the power-law corrected fit, Eq.\eqref{corr}. The empty circles are the simulation data and the vertical gray bars indicate the position of the BP. The red dashed line in the right panel is the total VDOS corrected by a Debye cutoff $\exp\left(-\omega^2/\omega_D^2\right)$, with $\omega_D \approx 13$. \textbf{(Bottom):} the black symbols are the data from simulations. The solid lines indicate the exponential fit and the dashed ones the power-law corrected ones.}
    \label{r1}
\end{figure}

We start by considering the simulation data for 2D and 3D  zero temperature glasses presented in Refs. \cite{Hu2022,PhysRevResearch.5.023055}. In Fig.\ref{r1}, we present two explicit examples performed using the data for the 2D power law model (2DPL) in \cite{Hu2022} and the 3D inverse power law model (3DIPL)  in \cite{PhysRevResearch.5.023055}. By plausibly assuming that the average velocity of the stringlet propagation coincides with the speed of transverse sound, we obtain a prediction for the BP of the 2D glass given by $\omega_{BP}^{th} \approx 1.344$ which has to be compared with the reported simulation value $\omega_{BP}^{\mathrm{sim}}=1.41$. The difference between the two is less than $5\,\%$. 

For the 3D system, the agreement is even superior, showing $\approx 1 \,\%$ uncertainty. The predictions of the theoretical model and the compared simulation values are reported in Table \ref{tab1}.
\begin{table}[!t]
\caption{The theoretical prediction versus the simulation value for the BP frequency using the exponential fit, Eq.\eqref{di}. The uncertainty $\Upsilon$ is defined as $\Upsilon= 2 |\omega_{BP}^{th}-\omega_{BP}^{\mathrm{sim}}|/(\omega_{BP}^{th}+\omega_{BP}^{\mathrm{sim}})$. All quantities presented here are in reduced Lennard-Jones units. The values of the BP frequency are those reported in Refs. \cite{PhysRevResearch.5.023055,Hu2022}.}%
\begin{tabular*}{\columnwidth}{@{\extracolsep\fill}llllll@{\extracolsep\fill}}
\toprule
 & $\lambda$  & $v_T$  & $\omega_{BP}^{th}$  & $\omega_{BP}^{\mathrm{sim}}$ & $\Upsilon$ \\
\midrule
3DIPL \cite{PhysRevResearch.5.023055} & $1.335$ & $4.23$ & $2.487$ & $2.46$ & $1\,\%$\\
2DPL \cite{Hu2022} & $3.717$ & $4.78$ & $1.344$ & $1.41$ & $4.7\,\%$\\
\botrule
\end{tabular*}
\label{tab1}
\end{table}
The estimates from the theoretical model are in remarkable agreement, \textit{i.e.}, agreement to within $5\,\%$ uncertainty, with the value for the BP frequency obtained from simulations and reported in Refs. \cite{PhysRevResearch.5.023055,Hu2022}. It is emphasized that, once the distribution of length is extracted from the simulation data and the value of the speed of transverse sound is taken from the simulation dispersion relation, the prediction of the BP frequency involves no free parameters.

In order to test further our model, we have extracted the simulation data from \cite{Hu2022,PhysRevResearch.5.023055} and compared them directly to the total VDOS with our predictions. In order to do so, we added a Debye contribution to the stringlet DOS:
\begin{equation}
    g_{\mathrm{total}}(\omega)=g_{\text{Debye}}(\omega)+g_s(\omega)=\mathcal{A}\,\omega^2+\mathcal{N}_s\,\frac{  \,e^{-\frac{\pi   v}{\lambda\,\omega}}}{\omega^2},
\end{equation}
where the Debye level $\mathcal{A}$ is fixed using the numerical data, and where we use the normalization constant $\mathcal{N}_s$ in the stringlet density of states, Eq.\eqref{final2}, as a fitting parameter. Additionally, as evident from the figures in the bottom panel of Fig.\ref{r1}, a more careful analysis reveals that the fit to an exponential becomes rather uncertain for longer strings which are relatively rare in the simulation sampling, \textit{i.e.}, $l>15$ in reduced LJ units. This was already observed in \cite{10.1063/1.4878502}, where a combination of a power-law with an exponential was considered and shown to fit the data better. In order to investigate this point further, we have performed a second analysis in which the stringlet length distribution $p(l)$ is not fitted with the exponential form, Eq.\eqref{di}, but rather with a power-law corrected expression as in Eq.\eqref{corr}.

We compare our analytic expression for the full reduced density of sates to our analytic model prediction in the top panels of Fig.\ref{r1}. The simulation data are shown with black symbols, the theoretical prediction based on the exponential distribution, Eq.\eqref{di}, is presented with solid black lines, the theoretical prediction based on the power-law corrected distribution Eq.\eqref{corr} by the dashed black lines, and finally the gray vertical bar indicates the position of the BP. In the two-dimensional model (2DPL) the comparison between the simulation data and our theoretical prediction is rather good, specially for the exponential fit. The fitting corrected by a power-law term seems to give a slightly better result for frequencies below the BP (corresponding to large $l$ in the distribution function $p(l)$), but a worse result for frequencies around and above the BP. In the three dimensional system, the comparison between the model and the simulations is good below $\omega \approx 4$ (in reduced units), but it becomes poor for large frequencies, much above the BP scale. In particular, the simulation data seem to decay much faster than our theoretical expectations. This discrepancy can be easily understood by noticing that our theoretical prediction does not include any UV cutoff to take into account the microscopic scale of the system. By adding a phenomenological Debye cutoff to the total density of states, $\exp \left( -\omega^2/\omega_D^2\right)$, this discrepancy immediately vanishes as shown by the red dashed curve in the top right panel of Fig.\ref{r1}. Finally, we notice that this discrepancy does not appear in the 2D model because the comparison does not extend to large enough frequencies.

\section{Damped dynamics in cooled liquids above the glass transition}
\begin{figure}
    \centering
    
    \includegraphics[width=0.52\linewidth]{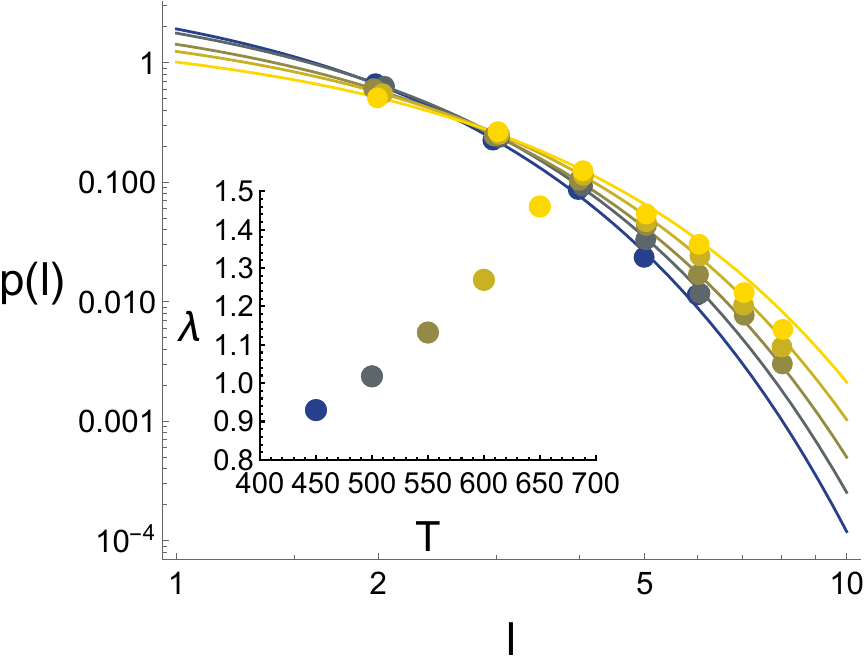}\quad
    \includegraphics[width=0.43\linewidth]{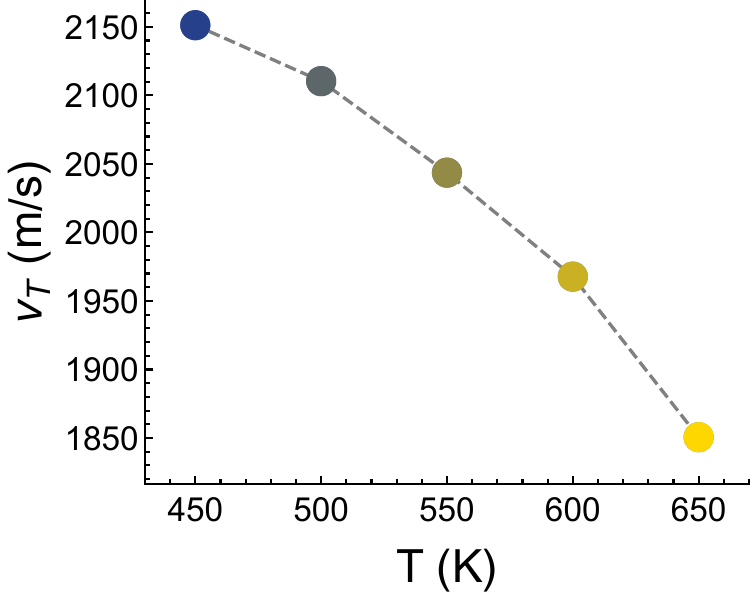}
    \caption{\textbf{Left panel: }The distribution of the stringlet lengths for the finite temperature Al-Sm system in \cite{10.1063/5.0039162} and the fits to the exponential form, Eq.\eqref{di}. The inset shows the dependence of the parameter $\lambda$ as a function of temperature. \textbf{Right panel: }The speed of transverse sound, extracted from the simulation data using Eq.\eqref{spee}, as a function of temperature. The values of $\lambda$ and $v_T$ are reported explicitly in Table \ref{tab2}.}
    \label{fig:s1}
\end{figure}

Inspired by these positive outcomes in the comparisons of the stringlet model to simulations deep in the glass state, we next consider simulations for a simulated Al-Sm metallic glass-forming material in a temperature range above $T_g$ discussed previously in \cite{10.1063/5.0039162} (see also \cite{Zhang2021}). By using the simulation values reported in Table \ref{tab2}, see also Fig.\ref{fig:s1}, we are again able to provide a parameter-free theoretical prediction for the BP frequency in the limit of zero damping, which is shown in the last column of the same table. 

\begin{table}[!t]
\caption{The parameters for the Al-Sm system at finite temperature \cite{10.1063/5.0039162}. $\lambda$ is reported in nanometers, the velocity in m/s and the frequencies in meV. The last column indicates the theoretical prediction for zero stringlet damping which has to be compared with the previous column reporting the corresponding simulation values. The values of the BP frequency correspond to those reported in Ref.\cite{10.1063/5.0039162}.}%
\begin{tabular*}{\columnwidth}{@{\extracolsep\fill}lllll@{\extracolsep\fill}}
\toprule
Al-Sm \cite{10.1063/5.0039162}  & $\lambda$  & $v_T$   & $\omega_{BP}^{\mathrm{sim}}$ &  $\omega_{BP}^{th}$ $(\gamma=0)$\\
\midrule
$T=450$K & $0.930$ & $2151$ & $2.681$ & $7.52$\\
$T=500$K  & $1.018$ & $2110$ &  $2.558$ & $6.74$\\
$T=550$K  & $1.132$ & $2043$ &  $2.348$ & $5.87$\\
$T=600$K  & $1.269$ & $1968$ &  $2.329$ & $5.04$\\
$T=650$K & $1.460$ & $1851$ &  $2.226$ &  $4.12$\\
\botrule
\end{tabular*}
\label{tab2}
\end{table}
The stringlet length distribution and the parameter $\lambda=\langle l\rangle-1$ are shown for different temperatures in the left panel of Fig.\ref{fig:s1}. Moreover, we obtained the values for the density $\rho$ and the shear modulus $G$ from \cite{10.1063/5.0039162,Zhang2021} and estimated the transverse speed of sound using the well-known relation:
\begin{equation}\label{spee}
    v_T^2=\frac{G}{\rho}.
\end{equation}
The data for $v_T$ required for our purposes are summarized in Table \ref{tab2} and shown in the right panel of Fig.\ref{fig:s1}. 

\begin{figure}[ht!]
    \centering
    \includegraphics[width=0.42\linewidth]{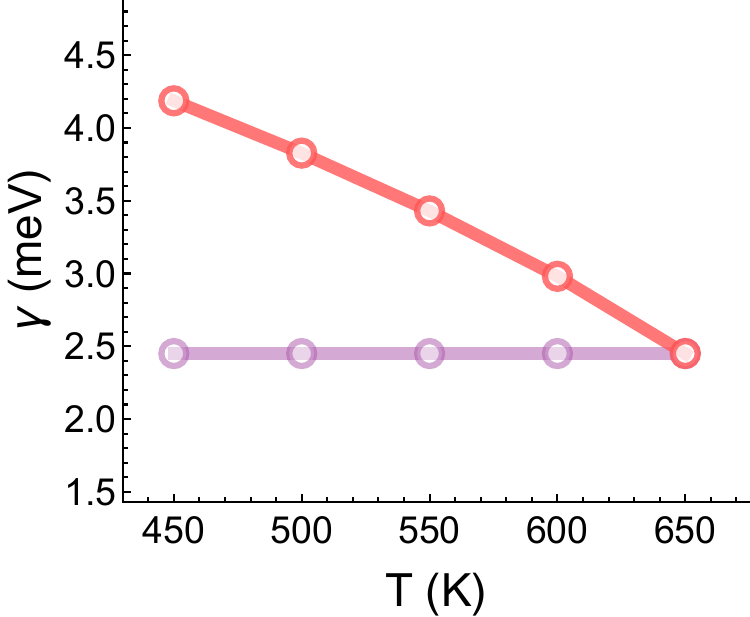}\quad
    \includegraphics[width=0.53\linewidth]{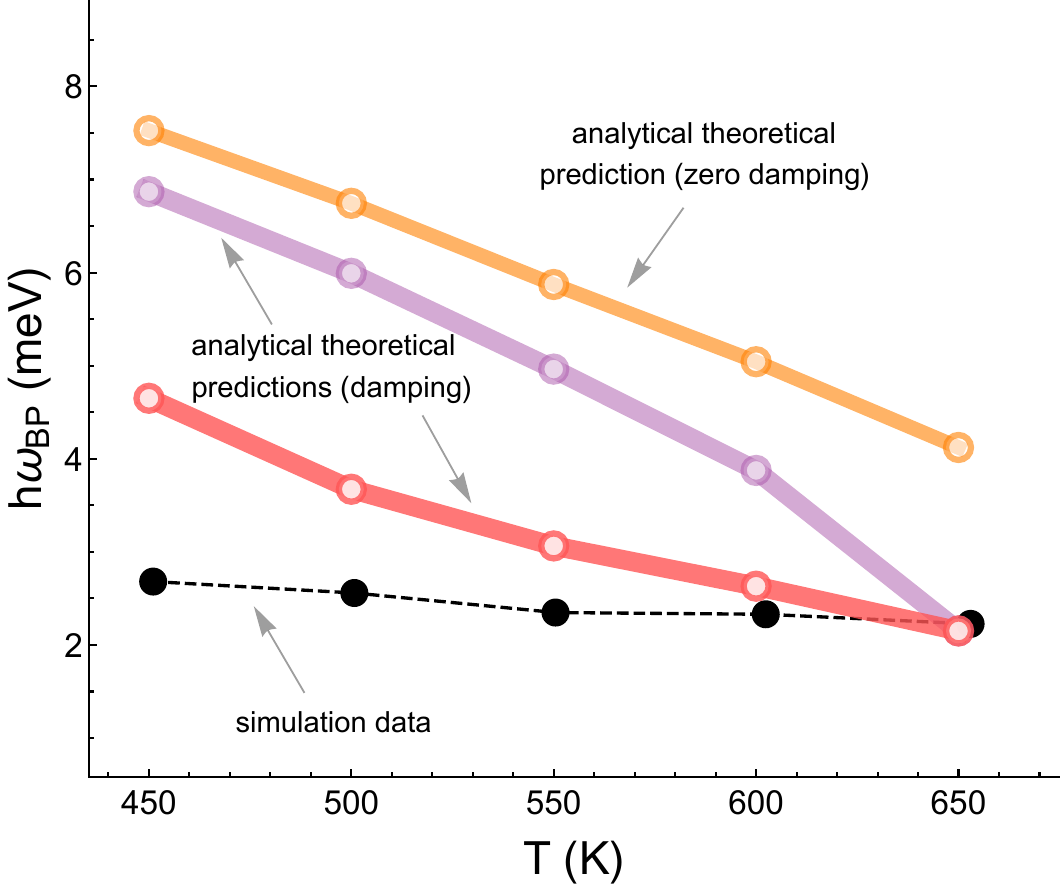}
    \caption{\textbf{Left panel: }The phenomenological damping parameter used in the theoretical predictions at high temperature, above $T_g$, for the Al-Sm system \cite{10.1063/5.0039162}. The colors match those used in the bottom panel. \textbf{Right panel: }The comparison between the theoretical predictions and the simulation results for the finite temperature Al-Sm system in \cite{10.1063/5.0039162}. The black symbols are the simulation data. The orange line indicates the theoretical results in absence of stringlet damping, $\gamma=0$. The pink and red line indicate the theoretical results in presence of damping, as explained in the text.}
    \label{fig:s2}
\end{figure}
Some insight in the damping parameter can be obtained by adopting idealized assumptions about its temperature dependence and comparing to the observed behavior of how the boson peak varies with temperature. As a first crude approximation, we take $\gamma$ to be a constant (pink curve in the left panel of Fig.\ref{fig:s2}) and not excessively large in magnitude, \textit{i.e.}, $\gamma$ smaller than $\omega_0$ in Eq.\eqref{ab}. We see that this option does not allow for a reasonable estimate of the BP frequency; see the pink curve in the bottom panel of Fig.\ref{fig:s2}. As it might have been anticipated, damping must be temperature dependent and relatively large in magnitude at temperatures well above $T_g$. Next, we consider for the sake of argument a temperature dependent $\gamma$ that is larger in amplitude but that decreases with temperature (see red curve in the left panel of Fig.\ref{fig:s2}), leads to a better accord between the stringlet model and simulation estimates of the boson peak, see red line in the bottom panel of Fig.\ref{fig:s2}, but the agreement is still inadequate. Moreover, we find it hard to envision a microscopic explanation for why the damping should decrease with temperature. This is clearly a non-linear dynamics many-body phenomenon. A damping that is larger in amplitude and that grows with temperature would then appear to be the choice for the variation of $\gamma$ but it poses an immediate difficulty. Heuristically, this problem has a simple origin. A large damping is necessary to soften the stringlet frequency and consequently the predicted value for $\omega_{BP}$, which, in absence of damping, is too large compared to the simulation observations. Nevertheless, a very large damping in Eq.\eqref{ab} clearly makes the stringlet frequency uncertain when $\gamma$, along with the stringlet length and the transverse sound velocity, all depend on temperature. Eventually, a too large damping would make the stringlet frequency in Eq.\eqref{ab} complex valued, which is obviously an issue within the model considered. We leave the task of more quantitatively understanding the temperature dependence of the boson peak (and its intensity) above $T_g$ for future work.

One qualitative finding is evident from Table \ref{tab2} (see also Fig.\ref{fig:s2}). Simulation observations cannot be described by a theory neglecting stringlet damping. The theoretical predictions neglecting damping overestimate the observed values of $\omega_{BP}$ by more than a factor of two. The reason for this deviation is easy to understand. As opposed to the simulations of Hu and Tanaka \cite{Hu2022,PhysRevResearch.5.023055}, which were performed in the limit of $T=0$ where stringlet damping can be reasonably neglected, the simulations for the Al-Sm metallic glass forming material \cite{10.1063/5.0039162} were performed at relatively high temperatures corresponding to a highly cooled glass-forming metallic glass liquid \cite{10.1063/5.0039162}. At such temperatures, it is unreasonable to assume $\gamma=0$ even as a rough approximation. The treatment of the boson peak at higher temperatures is complicated by the fact that the stringlet length, the transverse sound velocity and the extent of damping described by $\gamma$ can all be expected to vary appreciably with temperature (see Figs. \ref{fig:s1} and \ref{fig:s2} for the temperature variation observed in the Al-Sm metallic glass-forming liquid). Nevertheless, as we will see below, we can still gain important physical insights into the variation of the density of states and the boson peak with temperature by considering the stringlet model with finite damping $\gamma$ and by considering a simplified description for it.

Before continuing, let us emphasize that our analysis does not take into account the existence of hydrodynamically-defined ``defects'', such as the often discussed four-leaf quadrupole defects \cite{10.1063/5.0069477}. Nevertheless, as argued in Ref.\cite{Hu2022}, these relatively large scale structures apparently correspond to a frequency range much below the boson peak, and thus not germane to the present analysis. In particular, we do not expect this type of defects to have a significant impact in the frequency range near where the boson peak is exhibited in the reduced density of states. It is nevertheless plausible that these additional defects might contribute to the stringlet damping, especially at elevated temperatures and/or large frequencies, and it is also highly likely that these defects are highly significant in connection with understanding plastic deformation in glassy materials. We certainly do not mean to imply these defects are not important in glass-forming materials. Finally, we note that the existence of quadrupole defects defined in relation to elastic distortions in the continuum field describing the material provides a well defined example of how defects might arise without any crystalline structure necessarily existing in the material in order to define a defect. This would seem to imply that a similar non-crystalline based definition might apply to stringlets so that these structures might genuinely be termed ``defects”. This possibility remains to be established.

First, for 3D systems, the stringlet model with finite $\gamma$ formally predicts an interesting feature. In particular, using the stringlet reduced VDOS presented in Eq.\eqref{final2}, one can derive the low-frequency behavior of the stringlet VDOS as,
\begin{equation}\label{div}
    \frac{g_s(\omega)}{\omega^2}=\frac{\mathcal{N}_s}{\gamma^2}\frac{1}{\omega}+\dots
\end{equation}

Note also that Eq.\eqref{div} implies that the VDOS is linear in frequency at low energy in the high temperature liquid state so that our stringlet model exhibits the correct physical behavior in both the low temperature solid and the high temperature liquid states. This behavior is consistent with the experimental observations in classical liquids \cite{PhysRevLett.63.2381,dehong2022,jin2023dissecting} where it has been rationalized based on the concept of unstable modes \cite{Keyes1997,doi:10.1073/pnas.2022303118}. Interestingly, string-like excitations have been also discussed in glass-forming liquids at low temperatures \cite{PhysRevLett.80.2338,10.1063/1.453836,10.1063/1.4878502}, so that it would be important to study further the BP in relation to stringlet dynamics by heating such cooled liquids to watch how the boson peak evolves in shape and eventually disappears. Notably, this disappearance of the boson peak upon heating is well-established experimentally \cite{PhysRevE.83.061508,PhysRevLett.104.067402,PhysRevLett.92.245508,Zanatta2011-ZANTEO-3,10.1063/1.2360275,González-Jiménez2023}. Similarities to the occurrence of glassy anomalies in materials undergoing commensurate-incommensurate transitions \cite{PhysRevLett.93.245902,PhysRevLett.114.195502,PhysRevLett.76.2334,jiang2023glassy,BILJAKOVIC20121741} and we suggest that the pursuit of this ``analogy" might offer potential insights into the boson peak in glass-forming liquids. We next tentatively explore the extension to higher temperatures where we may expect damping of the stringlets.\\

We start our discussion about the predictions of our stringlet model at elevated temperatures by noting that an increase of the average stringlet length $\langle l\rangle$ with temperature is not sufficient to reproduce the complete phenomenology of the ``softening'' of the BP, and the simultaneous emergence of a linear in $\omega$ VDOS as appropriate for liquids \cite{dehong2022}. At the outset, we admit that the observed softening of the BP upon heating might involve multiple mechanisms, among which the growth of the average stringlet length $\langle l\rangle=\lambda+1$, the decrease of the velocity $v$, or the increase of the stringlet damping $\gamma$. From the simulation data, only the variation of $\lambda$ and $v$ with $T$ is known, but we cannot currently directly access information about the dependence of $\gamma$ on $T$. Thus, we cannot fully resolve how each mechanism contributes to the BP within the frame of our model without introducing a model of $\gamma(T)$. We introduce a tentative model of $\gamma(T)$ below to  address this problem.

\begin{figure}[ht!]
    \centering
    \includegraphics[width=0.49\linewidth]{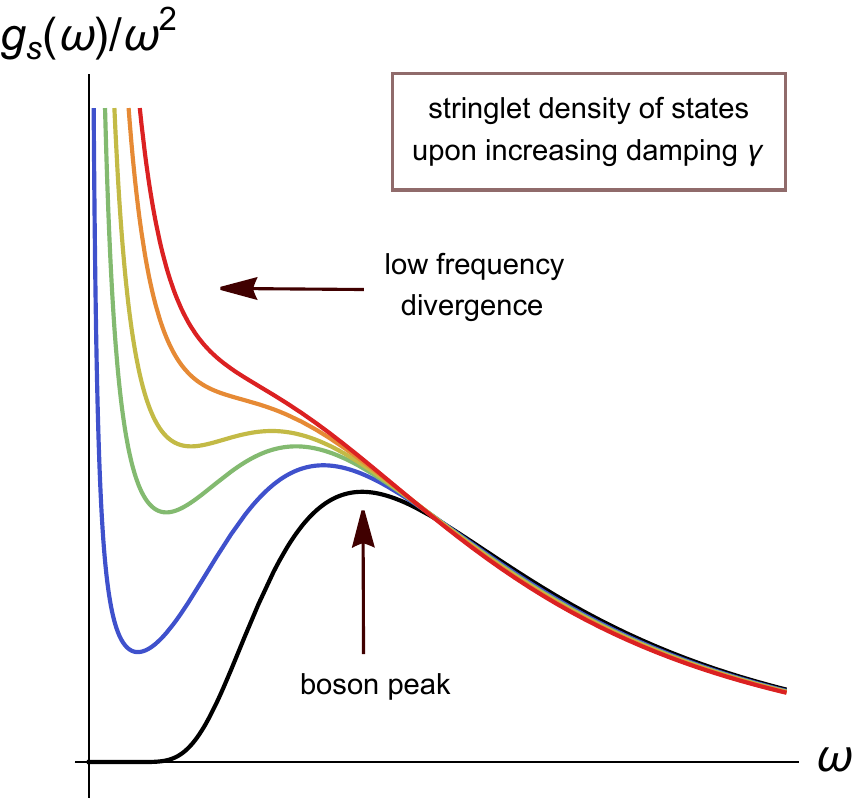}\quad
\includegraphics[width=0.47\linewidth]{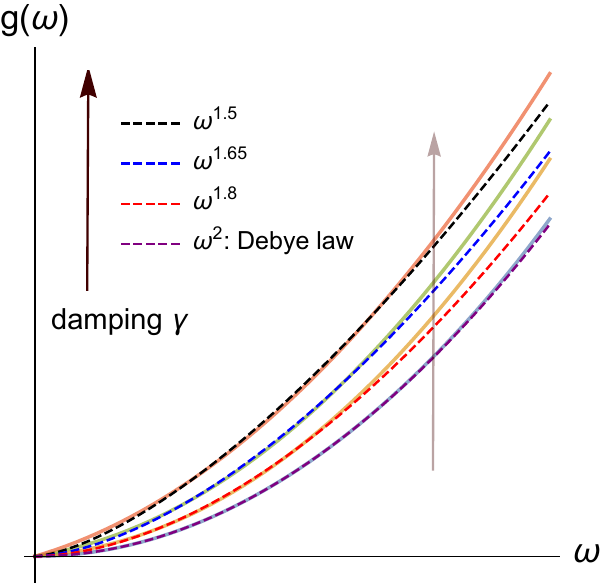}
    \caption{\textbf{Left panel: } The stringlet vibrational density of states normalized by Debye law $\omega^2$ from Eq.\eqref{final}. Here, we have fixed $\lambda=1$, $\mathcal{N}_s=2\pi$ and we move the damping $\gamma \in [10^{-4},0.25]$ from black to red. The arrows indicate the location of the BP and of the low-frequency $1/\omega$ divergence discussed in Eq.\eqref{div}. \textbf{Right panel: } The apparent power law emerging at low frequency in presence of damping. In this figure, we have fixed $\lambda=1$, $\mathcal{N}_s=2\pi$ and dialed the damping $\gamma \in [0,1]$. A Debye term $g_{\mathrm{Debye}}(\omega)=\omega^2$ has been added on top of the stringlet contribution, Eq.\eqref{final}.}
    \label{fig:4}
\end{figure}
Additionally, the combination of a Debye term, proportional to $\omega^2$, and a stringlet contribution with finite damping produces an interesting effect in the total VDOS. In particular, it induces an ``apparent power-law'' behavior at low frequency,
\begin{equation}
    g(\omega) \propto \omega^\zeta \qquad \text{with}\qquad 1<\zeta<2\,
\end{equation}
which is explicitly demonstrated in the bottom panel of Fig.\ref{fig:4}. The apparent power law $\zeta$ gradually moves from the Debye value, $\zeta=2$, for small damping towards a linear behavior, $\zeta=1$, for very large damping. As a consequence, for a reasonable value of damping, one expects an apparent power-law in between these two values. A power $\zeta \approx 4/3$ has been observed in superionic UO$_2$ heated crystal \cite{10.1063/1.5091042} and a similar power has also been found experimentally in neutron scattering measurements on a Cu$_{46}$Zr$_{54}$ metallic glass near room temperature \cite{Suck_1980}. It is tempting to rationalize such an intermediate power-law frequency scaling of $g(\omega)$ using the argument above and ultimately attribute its origin to the strong stringlet damping emerging at high temperatures. Unfortunately, we are not aware of a systematic simulation study of this effective power-law in a wide range of temperatures, which would be necessary to verify our hypothesis.

Interestingly, an expression for the critical damping at which the BP disappears in the stringlet model of the BP can be derived analytically. For simplicity, we do this only for a 3D system. The computation for the 2D system can be easily performed following similar steps.

Let us start by the reduced stringlet density of states in three dimensions derived in Eq.\eqref{final2},
\begin{equation}\label{rere}
   \frac{g_s(\omega)}{\omega^2}= \mathcal{N}_s\,\frac{ e^{-\frac{\pi  \,v}{\lambda  \sqrt{\gamma ^2+\omega^2}}}}{\omega
   \left(\gamma ^2+\omega^2\right)^{3/2}}.
\end{equation}
By defining $\omega^2 \equiv y$, the maximum of such a function, which corresponds to the location of the BP, is given by the solution of the following equation
\begin{equation}
    \pi \, v \,y \sqrt{\gamma ^2+y}-\lambda  \left(\gamma ^4+4 y^2+5 \gamma ^2 y\right)\,=\,0.
\end{equation}
Despite the solution to this equation is complicated, one can easily derive that such a solution involves the square root of the following quantity:
\begin{equation}
   \mathcal{F}(v,\lambda,\gamma)\equiv -432 \gamma ^4 \lambda ^4-4 \pi ^4
   v^4+207 \pi ^2 \gamma ^2 \lambda ^2 v^2\,.
\end{equation}
Whenever $\mathcal{F}(v,\lambda,\gamma)<0$, then, the solution becomes complex and therefore the function in Eq.\eqref{rere} does not show a maximum for real values of the frequency $\omega$. In order to have the solution real, and therefore a well-defined boson peak, the stringlet damping has to obey the following inequality
\begin{equation}
    \gamma < 0.466 \frac{v}{\lambda}.
\end{equation}
This depends explicitly on the value of the stringlet speed and average length $\langle l\rangle=\lambda+1$. However, we can obtain a universal result by considering a dimensionless stringlet damping $\tilde \gamma$ defined as the ratio between the stringlet damping and the BP frequency at $\gamma=0$, or equivalently at zero/low temperature,
\begin{equation}
    \omega^{(0)}_{BP}\equiv \frac{\pi}{4}\,\frac{v}{\lambda}.
\end{equation}
In doing so, we obtain a universal prediction for the disappearance of the BP within this simple model. Whenever the dimensionless damping reaches the critical value:
\begin{equation}
    \tilde \gamma^{\mathlarger{\mathlarger{*}}} \equiv  \left(\frac{\gamma}{\omega^{(0)}_{BP}}\right)^{\mathlarger{\mathlarger{*}}} = 0.568\,,
\end{equation}
the BP then disappears as it is absorbed in the low-frequency $1/\omega$ tail shown in the left panel of Fig.\ref{fig:4}.
This is a prediction of our theoretical model which can expressed as a function of temperature. Notably, this reduced damping parameter is taken to have a limited variation between $)$, corresponding to the zero temperature limit where no damping exists to a value  of $1$ appropriate to an ideal gas limit in which all normal modes are unstable, \textit{i.e.}, ``damped". This definitions avoids unphysical variations of $\gamma$ that can arise formally in our model if $\gamma$ is taken to be just a free parameter. Interestingly, if we think of $\omega_{BP}$ as the average frequency of vibration of the stringlets, this critical value is close to the onset of overdamped dynamics in which the oscillations of the stringlets are rapidly damped and decay exponentially.

\section{A simple physical model for a dimensionless stringlet damping}
Here, we introduce a simple physical model of the dimensionless stringlet damping parameter $\tilde \gamma (T)$, defined as the physical damping reduced by the zero temperature ($\gamma=0$) boson peak frequency, and estimate this quantity in terms of a readily observable property in both glass-forming and crystalline materials at elevated temperatures that quantifies the anharmonicity in interparticle interactions. Our model is based on the presumption that the apparently universal initial decay of the intermediate scattering function in both crystalline and glass materials arises from an increasing fraction of unstable normal modes that facilitate the fast $\beta$-relaxation process observed in condensed materials at low temperatures. The universality of fast dynamics in condensed materials at low temperatures makes it natural to consider the fast dynamics of crystalline and non-crystalline solid materials in a unified way.

These unstable modes in heated crystalline and glass materials can be calculated from a direct instantaneous normal mode analysis where it is found that these general classes of materials appear rather similar from this perspective \cite{10.1063/1.464106}. Clapa et al. \cite{10.1063/1.3701564} have made the interesting observation that the ratio of the fraction of localized unstable modes to stable localized modes in a model binary Lennard-Jones glass-forming material is rather similar to the ratio of all unstable modes to stable modes estimated from systematic INM analysis of localized and delocalized stable and unstable normal modes over a wide range of temperatures. These findings are relevant to the present analysis since Zhang et al. \cite{10.1063/5.0039162} suggested the identification of the stringlets with localized stable modes in the material and that the the relaxational contribution to fast dynamics arises from the loss of stability of these same modes, where the relaxation process involves irreversible particle exchange motion rather than the reversible periodic motion required of stable modes. String-like collective motion on a ps taking the form of both reversible and reversible motion was directly observed by Zhang et al., supporting this physical interpretation of both the boson peak and the fast relaxation dynamics in terms of stable and unstable modes taking the geometrical expression of string-like displacement motion. It is noted that all modes determined from an instantaneous normal mode (INM) analysis are stable at zero temperature and this picture of fast dynamics relaxation simply involves the suggestion that the localized stable modes found at zero temperature bifurcate into stable and unstable localized modes fractions at elevated temperature. This situation is highly plausible because a similar bifurcation of the stable modes in the zero temperature limit into families of stable and unstable modes at finite temperature is a general finding of essentially any INM analysis of condensed materials in their solid and liquid states. Mode damping at elevated temperatures is then a general phenomenon that can be expected to occur for both the localized stringlet modes, which we interpret as the ultimate physical origin of the boson peak, and structural relaxation on a ps timescale (``fast relaxation"), which we correspondingly attribute to unstable striglets.

Supporting this theoretical interpretation of the origin of both the boson peak and fast relaxation in cold condensed materials broadly, it has been repeatedly observed experimentally that a close inter-relationship exists between the ``vibrational contribution" of scattering identified with the boson peak and the ``relaxational contribution" to scattering intensity in the fast dynamics regime. This has been evidenced by the constancy of ratio of the intensities of depolarized and polarized light scattering, the polarization ratio at fixed temperature, as well as the light to excitation coupling coefficient $C(\omega)$ for both polarized and depolarized light scattering observations exhibiting the same anomalous near linear frequency dependence over a large frequency regime. Notably, this frequency scaling of $C(\omega)$ is consistent with the expectation of 1-dimensional string structures \cite{novikov1990spectrum,10.1063/5.0039162}. As noted previously, the fast dynamics relaxation time obtained from Brillouin scattering measurements has been found to scale inversely with the boson peak frequency \cite{PhysRevB.55.R14685}. These and closey related observations regarding the strong interrelationship between the boson peak and relaxational dynamics in the fast dynamics regime are summarized by Ngai et al. in Ref.\cite{10.1063/1.474889}. Finally, we point out that only the relative intensities of these elastic and inelastic scattering features, corresponding to the relative contributions of the vibrational and relaxational dynamics of the material, vary with temperature, suggesting a corresponding shift in the relative fraction of stable to unstable normal modes. Our mathematical model of strong ``damping", introduced below, is entirely consistent with this phenomenology and relates the temperature dependence of the relative fraction of these modes to the temperature dependence of the non-ergodicity parameter, quantifying the height of the intermediate scattering function at the end of the fast beta relaxation process.

This novel model of the fast dynamics of condensed materials relies on two important computational observations. (I) The observed string-like collective motion on a ps timescale completely dominates the magnitude of the mean particle displacement $\langle u^2 \rangle$ on the fast beta relaxation time because of the relative large displacements of the atoms involved in the stringlet motion \cite{10.1063/1.5009442}. Notably, $\langle u^2 \rangle$ does not describe a uniform increase in the amplitude of vibrational motion of the particles in the lattice as you might expect in a harmonic crystal. The anharmonic interparticle interactions of particles at finite temperature make materials rather generally ``dynamically heterogeneous” even on a ps timescale, and the string-like particle collective  exchange motion is closely tied to this phenomenon. (II) The magnitude of the non-ergodicity parameter of the self-intermediate scattering function on a ps timescale is a uniformly decaying function of $\langle u^2 \rangle$ in both crystalline and glass-forming materials \cite{10.1063/1.4878502,doi:10.1021/acs.macromol.3c00184}. This means that the non-ergodicity parameter encodes information about the relative stability of the vibrational modes responsible  for the boson peak, and thus the average propensity of these modes to be damped.

We next translate this qualitative physical picture into a \textit{quantitative} model by assuming that the change in the intensity of the normalized intermediate scattering function, which can be readily estimated from either the measurement or simulation, directly reflects the fraction of localized stable modes (stringlets) that have become damped at any given temperature. We tentatively assume that the dimensionless $\tilde \gamma(T)$, \textit{i.e.}, $\gamma$ in Eq. \eqref{ab} divided by the boson peak frequency at $\gamma=0$, or $T=0$, can be estimated from the amplitude of the fast beta relaxation process, which is simply $1$ minus the so-called non-ergodicity parameter describing the amplitude of the $\alpha$-relaxation process. This estimate of $\tilde \gamma$ has the particular advantage of being readily measured experimentally.

It is also helpful for our modeling that recent work has shown that the $T$ dependence of the non-ergodicity parameter can be apparently described by a near-universal function of the mean square particle displacement $\langle u^2\rangle$ at the fast beta relaxation time simply because the non-Gaussian parameter is inherently small in the fast dynamics regime of relaxation \cite{10.1063/1.5009442,doi:10.1021/acs.macromol.3c00184}. The model just described leads to a precise estimate of $\tilde \gamma(T)$ given by,
\begin{equation}\label{ss}
   \tilde  \gamma(T)= \begin{cases}
       1-e^{-7.6\, \langle u^2 \rangle/\sigma^2} \qquad T>T_o\\
        0\qquad \qquad \quad \quad \,\,\,\,\,\,\quad T<T_o
    \end{cases}
\end{equation}
where $\sigma$ the average interatomic distance estimated from the first peak position of the static structure factor. $T_o$ is the temperature at which $\langle u^2\rangle$ extrapolates to zero, indicating the temperature at which the glass-formation process ``ends'' (see Supplementary Information in \cite{Zhang2021} for more details). The constant $7.6$ in Eq.\eqref{ss} derives from the average interparticle distance of the atoms estimated from the peak of the static structure expressed in units of sigma and has been validated using MD simulations in \cite{xu2023parallel}. We notice that $\tilde \gamma(T)$ ranges between between $0$ and $1$, just as the ratio of the fraction of unstable localized and delocalized modes to their stable counterparts. In particular, the fraction of unstable modes is likewise known to increase with temperature as anharmonic interparticle interactions become progressively more prevalent \cite{Keyes1997}, and as more normal  modes corresponding become progressively unstable, \textit{i.e.}, ``damped". Finally, we note that the expression in Eq.\eqref{ss} applies quantitatively to polymer materials having a wide range of structures and a wide range of crystalline materials.

\begin{figure}[h]
    \centering
    \includegraphics[width=0.75\linewidth]{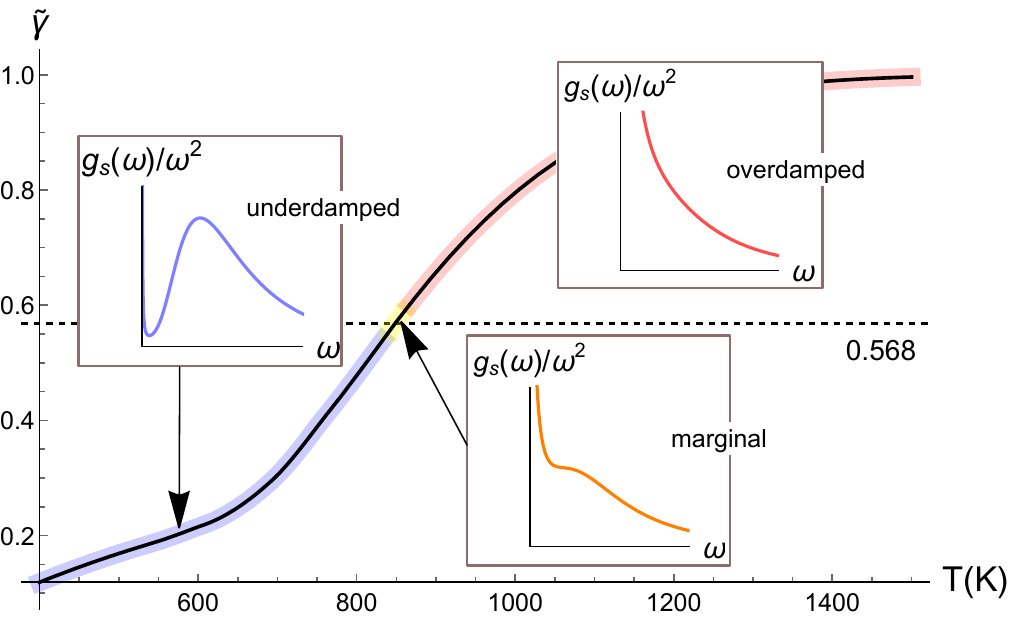} 
    \caption{Model estimate of the dimensionless stringlet damping $\tilde \gamma(T)$ as a function of temperature based on the identification of the intensity in the fast relaxation process as arising the damping of localized stable modes identified as being stringlets in previous simulations \cite{10.1063/5.0039162}. The horizontal dashed line indicates the critical damping value $\tilde \gamma ^{\mathlarger{\mathlarger{*}}}\approx 0.568$ which corresponds to the characteristic temperature $T^{\mathlarger{\mathlarger{*}}} \approx 848$ K.}
    \label{fig:new}
\end{figure}
\begin{figure}[h]
    \centering
    \includegraphics[width=0.7\linewidth]{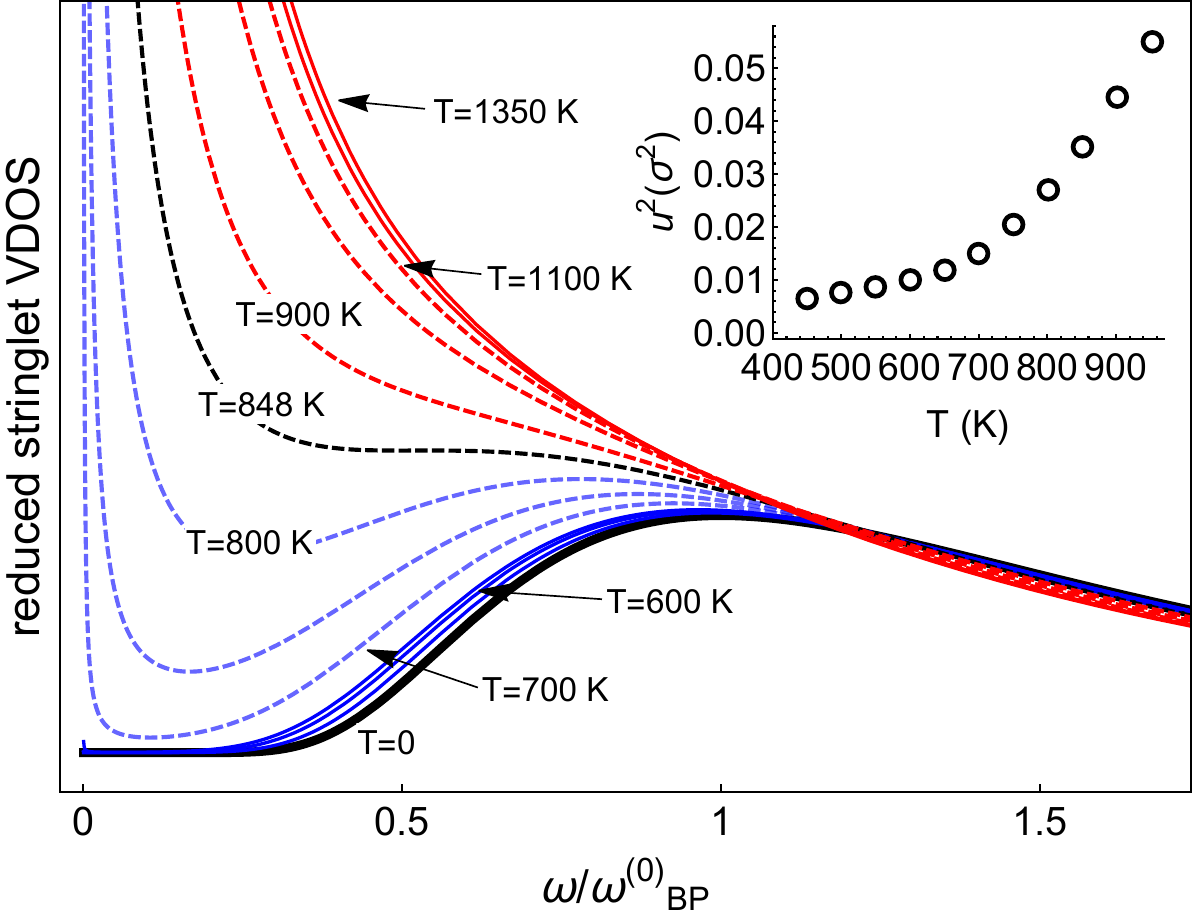} 
    \caption{The reduced stringlet density of states as a function of the frequency normalized by the zero temperature (or equivalently $\gamma=0$) BP frequency for different temperatures between $0$ K to $1350$ K (from blue to red). Solid lines are used to indicate the regimes in which the reduce DOS does not vary much with temperature. The filled black line is the result at zero temperature or $\gamma=0$. The dashed black line is the reduced stringlet DOS at $T^{\mathlarger{\mathlarger{*}}}\approx 848$ K at which the BP simply disappears. In the inset, the data for the mean squared displacement $\langle u^2\rangle$ (in units of $\sigma^2$) at the fast beta relaxation time \cite{10.1063/1.5009442} data for the Al-Sm metallic glass system are taken from \cite{Zhang2021}.}
    \label{fig:new2}
\end{figure}
In order to test this model and its predictions we have used the simulation data for the Al-Sm metallic glass system in \cite{10.1063/5.0039162,Zhang2021} for which the temperature dependence of the mean square displacement is known (see inset in Fig.\ref{fig:new2}). Moreover, we have used the fact that $\sigma \approx 2.8\, $Angstrom at $500$ K for this material and for simplicity we assume $\sigma$ is approximately independent of temperature. We plot our estimate of $\tilde \gamma(T)$ in Fig.\ref{fig:new}. We see that $\tilde \gamma(T)$ varies slowly below $\approx 650 $ K. It then strongly increases when the temperature becomes larger, between $650$ K and $1200$ K, and then plateaus at very high temperatures. Above $\approx 650$ K, $\langle u^2\rangle$ does not vary linearly as for particles subject to harmonic interparticle interaction, so the large and non-linear variation of $\langle u^2\rangle$ reflected in the strong increase of $\tilde \gamma$ provides clear evidence of large anharmonic interactions in the material where the boson peak is predicted by this model to become significantly damped.  In Fig.\ref{fig:new2}, we show the corresponding reduced stringlet density of states as a function of the frequency normalized by the $T=0$, or $\gamma=0$, BP frequency where the dimensionless damping $\tilde \gamma$ is modelled based on Eq.\eqref{ss}.

Evidently, the position of the BP approximately does not vary significantly with temperature inside the glass state (in Ref.\cite{10.1063/5.0039162}, the glass transition temperature was estimated to be close to $500$ K for the metallic glass under consideration) and it starts to exhibit appreciable softening only above this low temperature onset. As noted above, the BP peak disappears above a characteristic temperature $T^{\mathlarger{\mathlarger{*}}} \approx 848$ K in the AL-Sm glass-forming material, corresponding to the analytically derived critical damping condition discussed in the previous section. Above $T^{\mathlarger{\mathlarger{*}}}$, the stringlet excitations are predicted to be fully overdamped and the DOS then approaches the linear behavior $g_s(\omega)\sim \omega$ of a simple fluid. Notice that at elevated temperatures, above $\approx 1200$ K, the VDOS does not vary much as a simple consequence of the dimensionless damping $\tilde \gamma$ (Fig.\ref{fig:new}) approaching a constant value. All these features are in good qualitative agreement with experimental and simulation data. Let us emphasize that the scale at which strong temperature variations appear should relate to the scale at which strong deviations from the scaling $\omega_{BP} \sim G^{1/2}$ are observed, suggesting a possible common origin for these two phenomena.

We conclude with a brief discussion about the critical temperature $T^{\mathlarger{\mathlarger{*}}}\approx 848$. This scale is well above $T_g \approx 500$ K, but close to the onset temperature $T_A$ defining the onset of Arrhenius dynamics for glassy dynamics which has been independent estimated to be around $900$ K \cite{10.1063/5.0039162}. This temperature exactly corresponds to where the intermediate scattering function first develops a discernible multi-step decay, and at which the non-ergodicity parameter becomes measurable. The structural relaxation time typically becomes on the order of a ps at this characteristic temperature and non-Arrhenius relaxation, stretched exponential decay, etc., also first arise below this temperature. The possibility that all these phenomena are just downstream consequences of the stringlets acting on  ps timescale is an interesting possibility to consider in the future.
\section{Outlook} 
In this work, we showed that a simple theoretical model based on a distribution of 1D vibrating strings \cite{PhysRevB.91.094102} -- the stringlet model of the boson peak -- provides a good description of the BP frequency $\omega_{BP}$ for 2D and 3D amorphous systems in the glass state, where the damping of these excitations is reasonably be assumed to be negligible. In particular, this theoretical framework gives an analytical, and parameter-free, prediction for $\omega_{BP}$ that is in quantitative agreement with recent simulation data at $T=0$ \cite{Hu2022,PhysRevResearch.5.023055}. Moreover, the observed scaling of the BP frequency with the shear modulus in this type of zero temperature simulation, $\omega_{BP}\sim G^{1/2}$, follows as an analytic prediction from the stringlet model. 

We tentatively explored the stringlet model in the relatively high temperature regime of glass-forming liquids above $T_g$ where some damping of the stringlet modes can naturally be expected. In this temperature regime, we addressed the qualitative temperature dependence of the boson peak frequency and its disappearance at high temperatures, in connection with the dynamic correlation length introduced phenomenologically in previous work investigating boson peak measurements \cite{PhysRevE.83.061508,HONG2011351}. The softening of $\omega_{BP}$ upon heating is reasonably reproduced by the stringlet model and attributed mainly to the change of the stringlet length and transverse sound velocity with temperature. Our qualitative analysis suggests that the stringlet modes become completely overdamped in the liquid regime at sufficiently high temperatures. In association with this phenomenon, we observe the emergence of a density of states that changes to a linear variation with frequency rather than the quadratic law expected for the Debye theory of solids, a transition noted in recent studies attempting to define a density of states appropriate to materials in their liquid state \cite{jin2023dissecting,Keyes1997,doi:10.1073/pnas.2022303118,dehong2022}. We finally mention that recent studies \cite{10.1063/5.0039162,C2SM26789F,10.1063/1.5009442,Hu2022} of the boson peak frequency as a function of temperature show a nearly linear behavior of $\omega_{BP}$ with $G$, rather than $\propto G^{1/2}$, an observation that offers some clues about the variation of the boson peak in the temperature regime where the average stringlet length, the sound velocity and $\gamma$ are all varying in their own individual ways with temperature. In particular, Zhang et al. \cite{10.1063/5.0039162} indicated that since the increase of $\langle u^2 \rangle$ and the occurrence of fast beta relaxation first becomes appreciable near the Vogel-Fulcher-Tammann temperature at which the $\alpha$-relaxation time extrapolates to infinity. Stringlet damping should also begin to arise near this same temperature. Rossi et al. \cite{caponi2011influence} have observed a deviation of the boson peak frequency from the Debye frequency at temperatures of about $200$ K below $T_g$, roughly consistent with the arguments of Zhang et al. that softening of the boson peak occur around $T_o$. These and other recent measurements of the pressure dependence of the boson peak \cite{PhysRevLett.99.055502,ahart2017pressure}, indicating a deviation from the simple scaling of the boson peak with the transverse sound velocity, offer some clues about this more complicated regime for temperatures somewhat below or above $T_g$. Further measurements along this line should be helpful in understanding the effect of excitation damping on the boson peak.

Our analysis supports the idea that the stringlets correspond to localized modes, which have long been suggested \cite{10.1063/5.0039162,Hu2022,PhysRevResearch.5.023055} to be responsible for the BP anomaly in amorphous systems,. However, the precise nature and origin of these modes remains an object of continued investigation. This identification raises questions about how these modes might be related to the multipolar four-leaf modes \cite{PhysRevLett.117.035501,doi:10.1073/pnas.1709015114} that have recently been discussed as being related to the the emergence of the boson peak. These structures would appear to be much larger than the stringlets suggesting to us the possibility that that stringlets may account for their substructure. This possibility also deserves further investigation.

In order to augment the predictive nature of the stringlet model, and to develop it into a full-fledged theory, clearly more work needs to be done. First, a theoretical model is required for the first-principle prediction of the stringlet length distribution $p(l)$, and in particular the total stringlet number proportional to $\mathcal{N}_s$, without resorting to simulation data. A promising avenue would be to combine Lund model \cite{PhysRevB.91.094102} with the the well-developed thermodynamic theory of equilibrium polymers discussed in Refs. \cite{10.1063/1.4878502,10.1063/1.2909195,10.1063/1.2356863}. Second, it would be interesting to directly estimate of the average string length $\langle l\rangle$ as a function of temperature, and compare it with independent estimates of the dynamic correlation length based on Eqs.\eqref{w1}-\eqref{w2}, to verify that that this length can be interrelated as cooperativity length, as proposed heuristically in some experimental studies \cite{PhysRevE.83.061508,HONG2011351}. Specifically, Eqs.\eqref{w1}-\eqref{w2} are equivalent to the ``standard'' relation taken to define the dynamic correlation length by Hong et al. \cite{PhysRevE.83.061508,HONG2011351} and others \cite{10.1063/1.2136878}, when the parameter $S$ of these experimental studies is taken to have the specific predicted value, $S=\pi/(d+1)$, with $d$ the number of spatial dimensions.

Additionally,  we have not considered in the present work the question of the intensity of the BP and how this relates to the characteristic features of the stringlets and their size distribution. The universality of the functional form suggests that the boson peak frequency and the height should be related. Using Eq.\eqref{final2} in the limit of negligible damping, $\gamma=0$, one can easily verify that the intensity of the stringlet contribution at the BP frequency is given by,
\begin{equation}\label{int}
  \frac{g_s(\omega_{BP})}{g_{Debye}^T(\omega_{BP})}\equiv  v_T^d \frac{g_s(\omega_{BP})}{\omega_{BP}^{d-1}} \propto \mathcal{N}_s \lambda^{d+1} ,
\end{equation}
where $g_{Debye}^T(\omega_{BP})$ is the Debye contribution arising from the transverse acoustic modes $\omega^{d-1}/v_T^d$, up to some neglected constant numerical factors. It is interesting to notice that, in this approximation, the intensity is given uniquely in terms of the parameters appearing in the stringlet size distribution $p(l)$. Also, it is immediate to verify that the intensity of the BP as in Eq.\eqref{int} grows upon heating since $\langle l\rangle=\lambda+1$ increases with $T$.

Interestingly, the stringlet dynamics shares many similarities with the behavior of defects and anharmonic modes in crystals \cite{PhysRevLett.68.974,TAKENO19881023} (See Ref.\cite{PhysRevB.105.014204} for a direct proof of the stringlet anharmonicity in relation to the BP in glasses.) and the role of weakly-dispersing soft optical modes, which has been identified in several instances as the origin for the BP anomaly in crystalline materials with no structural disorder \cite{PhysRevB.23.3886,PhysRevLett.76.2334,PhysRevLett.114.195502,PhysRevLett.93.245902,PhysRevB.99.024301,RevModPhys.86.669,Schliesser_2015,doi:10.1021/acs.jpclett.2c01224,Baggioli_2020}. A similarity of this kind is also reflected by the fact that the string model of Lund was inspired by Granato and L{\"u}cke string model of crystal plasticity \cite{granato1956theory}. A unifying picture based on defect dynamics would be highly desirable and these collective excitations are a candidate for explaining the roton-like excitations observed previously in both liquids and glasses \cite{PhysRevLett.49.1271,PhysRevLett.50.49,DESCHEPPER198429}, as reviewed in \cite{ASTRATH20063368}. Dispersion curves of this kind have also been suggested to be related to the BP anomaly in glasses \cite{KOVALENKO1990115,KRASNY199892} (see also \cite{PhysRevResearch.4.029001}), and might likewise be related to the stringlets.

Finally, our work revives the long-standing and vigorously debated question of whether the structures responsible for the BP are structural defects or dynamical excitations deriving from the anharmonicity of intermolecular interactions. The evidence would appear to favor the latter interpretation, explaining why the boson peak should also be expected in crystalline materials at elevated temperatures where anharmonicity in intermolecular interactions becomes prevalent as in cooled liquids. Recent simulations \cite{10.1063/1.5091042} have indeed confirmed that the boson peak also arises in crystalline materials at elevated temperatures, but below the melting temperature, where anharmonic intermolecular interactions start to become prevalent.


%
%

%

\begin{acknowledgments}
We would like to thank A.~Zaccone, J.~Zhang and M.~A.~Ramos for many discussions about the BP and the vibrational dynamics of amorphous solids. We are grateful to Y.~Wang. and W.~Xu for useful comments on a preliminary version of this draft. We thank P.~Jacobson and C.~Soles for helpful comments. We would like to thank H.~Zhang for the unpublished image in the right panel of Fig.\ref{fig:0} and several fruitful discussions. CJ and MB acknowledge the support of the Shanghai Municipal Science and Technology Major Project (Grant No.2019SHZDZX01). MB acknowledges the sponsorship from the Yangyang Development Fund.
\end{acknowledgments}
\section*{Data Availability Statement}
The datasets generated and analysed during the current study are available upon reasonable request by contacting the corresponding authors.
\appendix 
\section{Analogy between dislocations in crystals and stringlets}\label{app1}
Dislocations are extended “defect” structures that are prevalent in the modeling of the plasticity of crystalline materials. The model of “open” string-like defects utilized in the main text originates from a highly successful model of crystal plasticity introduced by Granato and L{\"u}cke \cite{granato1956theory} to describe the “string-like” segments of the dislocations between pinning points. The development of this classical dislocation model of deformed crystalline materials is reviewed in \cite{KHONIK2021157067}. 
In a more controversial work, Granato advocated that open string-like excitations in both crystalline materials and cooled liquids should correspond to extended defects involving ``generalized'' interstitial defects \cite{PhysRevLett.68.974}. Granato and his coworkers then advocated the radical idea that liquids should be viewed as crystalline materials populated by such defects. One of the many predictions of Granato's model \cite{PhysRevLett.68.974} strongly anticipates the results of the present paper. In particular, he argued that his interstitial defects should adopt a string like geometrical form and he ``re-purposed" his former dislocation segment theory to model the boson peak of glass-forming liquids. This led him to a prediction of the boson peak in the low temperature limit of the same form as Eq.(10), although his admittedly rough derivation differs in the predicted prefactor. Moreover, his model made a rough estimate of the boson peak frequency. Clearly, our more refined model has much in common with the Granato model of the boson peak and the correspondence becomes even greater if an identity is granted between the strings and his hypothetical interstitial defects. This is a possibility that we plan to study in the future.

While there is substantial experimental support for the ideas just described, it is currently unclear how one should interpret such defects in liquids (and amorphous systems in general) given there is no crystalline reference state that would allow for a discussion of well-defined structural defects, interstitial or otherwise. Nonetheless, both simulations and neutron scattering measurements have indicated that there are indeed “excitations” in the form of strings in the ps timescale of glass-forming liquids. Simulations show that such excitations also arise in model crystalline materials, although these structures cannot reasonably be identified as static dislocation loops or extended defects of composed of clustered interstitial defects \cite{Santos2018}.

It is our view that these string-like structures, that are notably observed in simulations of both crystalline and glass-forming materials, arise from the inherently anharmonic interactions in these materials, a phenomenon that is well-known in the Fermi-Ulam-Pasta-Tsingou model (\textit{e.g.}, \cite{campbell2004fresh,PhysRevE.73.036618}). In particular, exact and numerical solutions of the dynamics and thermodynamics of such non-linear lattice models reveals the presence of intrinsically localized modes (sometimes termed “breathers”) \cite{PEYRARD1998184,PEYRARD2000199}, deriving purely from the anharmonicity of the interparticle interactions. 
       
Importantly, the number of particles exhibiting this type of collective motion has been shown in the closely related $\phi^4$ lattice model to exhibit an exponential size distribution \cite{flach1994slow}, as for stringlets. Similar distributions of the number of particles in this type on non-linear expectation have been noted in the analysis of the Peyrard-Bishop model of duplex DNA \cite{PhysRevLett.62.2755}, another closely related anharmonic 1-dimensional lattice model. 
       
Our theoretical framework adopted to describe the density of states distribution of the stringlets is inspired mathematically by the Granato-L{\"u}cke dislocation segment theory, noted above, and by a thermodynamic theory of string formation at equilibrium introduced previously to describe the $T$ dependence and length distribution of strings \cite{10.1063/1.5009442}. Very similar statistical mechanical models have been introduced to describe the formation of worm-like micelles, the equilibrium polymerization of S, P and other elements, etc. 
       
Despite the success of our modeling effort aimed at fundamentally understanding the boson peak, we must admit that the fundamental cause of this type of “excitation” is still not sufficiently clear. It is notable that string-like excitations are prevalent in many field theories of phase transitions and other thermodynamic transitions \cite{copeland1991statistical}, such as the XY model \cite{PhysRevLett.57.1358} and the superfluid to fluid transition in He$_4$ whose critical properties are well-described  by this model, vortex melting in type II superconductors \cite{PhysRevB.57.3123}, and cosmological models \cite{zurek1985cosmological,PhysRevLett.80.908,PhysRevLett.81.3083}. We strongly believe that the formulation of such a gauge field theory of condensed matter along this line is required to address the fundamental origin of these dynamic excitations.
\bibliography{mybib}
\end{document}